\documentclass[conference]{IEEEtran}

\usepackage[table,xcdraw]{xcolor}
\usepackage{graphicx}
\usepackage{cite}
\usepackage{url}
\usepackage{hyperref}
\usepackage{balance}
\usepackage{svg}
\usepackage{xcolor}
\usepackage{subfig}
\usepackage{multirow}

\newcommand\revision[1]{{\color{blue}#1}}

\begin{document}

\title{A Case For Adaptive Deep Neural Networks\\in Edge Computing}



\author{\IEEEauthorblockN{Francis McNamee\IEEEauthorrefmark{1},
Schahram Dustadar\IEEEauthorrefmark{3},
Peter Kilpatrick\IEEEauthorrefmark{1},\\
Weisong Shi\IEEEauthorrefmark{4},
Ivor Spence\IEEEauthorrefmark{1}, and
Blesson Varghese\IEEEauthorrefmark{1}
}
\IEEEauthorblockA{\IEEEauthorrefmark{1}Queen's University Belfast, UK}
\IEEEauthorblockA{\IEEEauthorrefmark{3}TU Wien, Austria}
\IEEEauthorblockA{\IEEEauthorrefmark{4}Wayne State University, USA}
E-mail: fmcnamee04@qub.ac.uk,
dustdar@infosys.tuwien.ac.at, 
p.kilpatrick@qub.ac.uk,\\ 
weisong@wayne.edu, 
i.spence@qub.ac.uk, 
b.varghese@qub.ac.uk (Corresponding e-mail)
}

\maketitle

\thispagestyle{plain}
\pagestyle{plain}

\begin{abstract}
Edge computing offers an additional layer of compute infrastructure closer to the data source before raw data from privacy-sensitive and performance-critical applications is transferred to a cloud data center. Deep Neural Networks (DNNs) are one class of applications that are reported to benefit from collaboratively computing between the edge and the cloud. A DNN is partitioned such that specific layers of the DNN are deployed onto the edge and the cloud to meet performance and privacy objectives. However, there is limited understanding of: (a) whether and how evolving operational conditions (increased CPU and memory utilization at the edge or reduced data transfer rates between the edge and the cloud) affect the performance of already deployed DNNs, and (b) whether a new partition configuration is required to maximize performance. A DNN that adapts to changing operational conditions is referred to as an `adaptive DNN'. This paper investigates whether there is a case for adaptive DNNs in edge computing by considering three questions: (i) Are DNNs sensitive to operational conditions? (ii) How sensitive are DNNs to operational conditions? (iii) Do individual or a combination of operational conditions equally affect DNNs? (iv) Is DNN partitioning sensitive to hardware architectures on the cloud/edge? The exploration is carried out in the context of 8 pre-trained DNN models and the results presented are from analyzing nearly 8 million data points. The results highlight that network conditions affects DNN performance more than CPU or memory related operational conditions. Repartitioning is noted to provide a performance gain in a number of cases, but a specific trend was not noted in relation to its correlation to the underlying hardware architecture. Nonetheless, the need for adaptive DNNs is confirmed.
\end{abstract}

\IEEEpeerreviewmaketitle

\section{Introduction}
\label{sec:introduction}
Edge computing envisions that compute resources located or placed at the edge of the network, such as routers, gateways or dedicated micro data centers, may be used for running certain application services closer to the end-user device where data is generated~\cite{edgecomp-01, edgecomp-02, edgecomp-03, intro-01}. Processing data at the edge provides opportunities for making the application more responsive by reducing end-to-end latencies, performance efficient by reducing ingress bandwidth demand beyond the edge resource, and privacy-sensitive by selectively releasing data beyond the edge. 

Many performance-critical and privacy-sensitive applications are demonstrated to benefit from edge computing - for example, cognitive wearable assistance~\cite{cognitivewearable-01}, image and video analytics~\cite{couper}, connected and autonomous vehicles~\cite{cav-01} and privacy preserving denaturing~\cite{privacydenaturing-01}. These applications take advantage of the edge by distributing services of the application across the edge and the cloud. 

One reason the above applications lend themselves to take advantage of the edge is because they are underpinned by Deep Neural Networks (DNNs). A DNN is a sequence of multiple layers (each layer is a collection of neurons) that carry out functions, such as convolution, pooling or activation~\cite{dnn-01, dnn-02}.
Therefore, the layers of a DNN can be inherently distributed in a specific manner across the edge and the cloud to reduce inference times, reduce the volume of data transferred to the cloud from a sensor or end-device, or to not release sensitive data beyond the edge~\cite{edge-dnn-survey-01, edge-dnn-survey-02}. This is achieved by DNN partitioning - splitting the DNN into two sequential DNNs at a specific layer (the layer at which the DNN is partitioned is referred to as the partitioning point). The partitions can then be distributed across the cloud and the edge. For example, consider a DNN that has seven sequential layers and needs to be distributed across the cloud and edge. If the DNN is partitioned at the third layer, then the first three layers may execute on the edge and the remaining four layers on the cloud. 

DNN partitioning for performance efficiency in edge computing is an avenue that has been reported in the literature. There are multiple techniques for partitioning DNNs, such as using estimation-based~\cite{neurosurgeon, deepwear, musicalchair, couper}, structural modification-based~\cite{deepthings, modnn, dnncontainerized-01}, and measurement-based techniques~\cite{lavea}. These techniques identify an optimal partitioning point primarily based on the characteristics of the layers of a DNN and operational conditions, such as resource utilization or network conditions. 
From the literature reported above it is understood that distributed DNN execution has specific performance advantages. 

However, there is limited understanding of whether and how evolving operational conditions at the edge and hardware architectures affect the performance of the already deployed DNNs and raises an important question on whether a new partitioning configuration is required. If new system conditions affect performance, then the DNN will need to be repartitioned to maximize its performance under the new operational conditions or to suit new architectures. Such a DNN that adapts to operational conditions or hardware architectures is referred to in this paper as an `adaptive DNN' and this process is referred to as `adaptivity'. 

Therefore, the research reported in this paper sets out to investigate \textit{whether there is a case for adaptive DNNs in edge computing}. In doing so, the following four questions are considered: 

(\textbf{Q1}) \textit{Are DNNs sensitive to operational conditions?}
This question addresses whether operational conditions, such as CPU and memory stress (increased utilization) on edge resources or the network data transfer rate between the cloud and the edge, affect the performance of a distributed DNN. For example, it is important to understand if the operational conditions change, whether the initial DNN partitions are still the most performance efficient partitions under the new operational conditions. 

(\textbf{Q2}) \textit{How sensitive are DNNs to operational conditions?}
If Q1 is true, then the aim is to answer a second question, which is to identify how different would the partition configurations be (and difference in performance) given a change in the operational environment. In other words, whether new partitions will be significantly different (more or fewer layers) than the initial partitions. 

(\textbf{Q3}) \textit{Do individual or a combination of operational conditions equally affect DNNs?}
In this paper, the operational conditions are explored individually (only varying a single operational condition and by not explicitly influencing the other conditions) and in combination (multiple operational conditions are varied) to identify their sensitivity to DNN performance. For example, it is important to understand whether the partitioning point of a DNN changes when operational conditions change individually or in a combination. This paper will examine this question by controlling the change in operational conditions on a range of DNNs to examine the effect of operational conditions individually and in combination. 

(\textbf{Q4}) \textit{Is DNN partitioning sensitive to hardware architectures?}
In this paper, multiple hardware architectures, namely Intel and Arm processors are employed across the cloud and edge. Observations from these are valuable when deploying distributed DNNs in an architecture rich and heterogeneous distributed computing environment.  

The questions raised above have not been considered in the existing literature. Therefore it is essential that they are understood within the context of edge computing to further explore adaptive DNNs and maximally leverage the benefits of using edge resources. 
This paper presents a first such exploratory study to address the above three questions by developing a practical methodology to benchmark DNNs across cloud and edge resources. 
Experimental studies are carried out on 8 DNNs by examining different but realistic operational conditions for CPU stress, memory stress and network data transfer rates to address the above three questions.

The results presented are obtained from a cloud and edge lab-based experimental platform by analyzing nearly 8 million data points (the data obtained and the associated code for benchmarking are available for download\footnote{\url{https://github.com/qub-blesson/AdaptiveDNN}}). The key observation is that DNNs are sensitive to operational conditions. Both individual and a combination of operational conditions affect DNN performance. Network conditions have more impact on DNN performance than CPU stress and memory stress individually. Repartitioning can provide performance gains, but there are DNNs that are not significantly impacted. Operational conditions in combination affect DNN performance more than individual operational conditions considered in this paper. 
Thus, there is a case for repartitioning, i.e., the need for adaptive DNNs.

The remainder of this paper is organized as follows. 
Section~\ref{sec:background} provides a background to DNN partitioning and repartitioning. 
Section~\ref{sec:methodology} presents the methodology used in this paper to explore DNN adaptivity. 
Section~\ref{sec:experimentalstudies} provides a discussion on the results obtained from the experimental studies. 
Section~\ref{sec:relatedwork} presents related research in the area. 
Section~\ref{sec:conclusions} concludes this paper by presenting avenues for future research.

\section{Background}
\label{sec:background}

A DNN consists of an input layer, multiple hidden layers, and an output layer (each layer is a collection of neurons)~\cite{dnn-01,dnn-02}. There are different types of DNN layers, which include: (1)~Fully-connected layers, (2)~Convolution layers, (3)~Pooling layers, (4)~Activation layers, and (5) Softmax layers. 


\begin{table}[t]
\centering
\caption{Pre-trained DNN models available from Keras that is used in this paper; Type: S - sequential, N - non-sequential}
\label{tab:pretrainedmodels}
\begin{tabular}{@{}lcccc@{}}
\hline
\textbf{DNN Model}             & \textbf{Size (MB)}  & \textbf{Layers} & \textbf{Partition points} & \textbf{Type}\\
\hline
VGG16~\cite{vgg}            & 527   & 23     & 22           & S\\
VGG19~\cite{vgg}            & 548   & 26     & 25           & S\\
MobileNet~\cite{mobilenet}  & 16      & 93     & 92           & S\\
AlexNet~\cite{imagenet}     & 110     & 25     & 24           & S\\
DenseNet~\cite{densenet}    & 31      & 429    & 22           & N\\
ResNet50 \cite{resnet}      & 98     & 177    & 23           & N\\
ResNet50V2 \cite{resnet}    & 98     & 192    & 16           & N\\
LeNet~\cite{lenet}          & 7      & 11     & 10           & S \\
\hline
\end{tabular}
\end{table}

Eight DNNs are considered in this research and are shown in Table~\ref{tab:pretrainedmodels}. The size of a trained model and its corresponding weights, the number of layers in the DNN, the number of valid points for partitioning, and the type of the DNN is shown. These models are obtained from the Keras\footnote{\url{https://keras.io}} neural network library.

\begin{figure}[t]
\begin{center}
	\subfloat[Sequential DNN]
	{\label{fig:sequential}
	\includegraphics[width=0.4\textwidth]{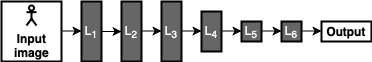}}
\hfill
	\subfloat[Non-sequential DNN]
	{\label{fig:nonsequential-1}
	\includegraphics[width=0.45\textwidth]{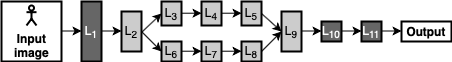}}
\hfill
	\subfloat[Non-sequential DNN after pre-processing]
	{\label{fig:nonsequential-2}
	\includegraphics[width=0.45\textwidth]{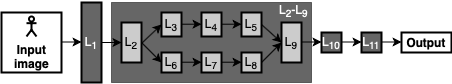}}
\end{center}
\caption{Sequential and non-sequential DNNs}
\label{fig:DNNtypes}
\end{figure}

There are two types of DNNs - sequential and non-sequential, represented as S and N, respectively in Table~\ref{tab:pretrainedmodels}.
In a sequential DNN the input of one layer is connected to the next in a linear manner (Figure~\ref{fig:sequential} shows an example with six layers). 
A non-sequential DNN on the other hand will have layers that may be connected to two or more layers (refer Figure~\ref{fig:nonsequential-1} for an example with 11 layers). Hence, there are multiple paths that connect the first and last layer of the DNN.

Distributing a sequential DNN across the cloud and the edge is straightforward. The DNN would need to be partitioned at a suitable layer that would yield maximum performance (for example, lowest end-to-end latency and/or least amount of data transferred from the edge to the cloud). 
For example, consider VGG19 shown in Table~\ref{tab:pretrainedmodels}, which has 26 layers and has 25 suitable partitioning points.  
Consider a DNN that has $N$ layers. If the $x^{th}$ layer is identified as the partitioning point, then the partition running on the edge would consist of a sequence of the first $x$ layers and the partition to be executed on the cloud would consist of $N-x$ layers. The output of the $x^{th}$ layer would need to be transferred from the edge to the cloud and provided as an input layer for the cloud partition. 

Distributing a non-sequential DNN requires additional pre-processing. This is to ensure that a parallel path in a DNN is not partitioned as it may lead to synchronization issues that will incur communication overheads~\cite{couper}. Partitioning is avoided on the parallel paths by grouping parallel layers as a single entity, referred to as a block of layers (refer to Figure~\ref{fig:nonsequential-1} and Figure~\ref{fig:nonsequential-2} for an example - Layers 2-9 are treated as a single entity). Therefore, the number of partitioning points is reduced. For example, in Table~\ref{tab:pretrainedmodels} ResNet50V2 has 192 layers, but with only 16 suitable partitioning points. 

The layer at which a DNN needs to be partitioned may depend on the characteristics of the DNN. For example, the layer at which the DNN is partitioned may be based on creating partitions that will result in the lowest execution time and the least volume of data that will be transferred between the edge and the cloud. Such a partitioning approach based on the DNN characteristics will create ideal DNN partitions.  

However, would an ideal partition be the most suitable for a given set of operational conditions, such as utilization of edge resources or network state between the cloud and edge? In addition, if an ideal partition were deployed and the operational conditions changed, would a more context-driven partition improve the overall performance of the DNN? 
For example, the edge may execute multiple workloads, resulting in increasing CPU utilization, which may affect the edge partition running on the network. Alternatively, the network between the edge and the cloud may be congested resulting in sub-optimal performance due to communication overheads. In such cases, the DNN performance may be sensitive to the operational conditions. As already highlighted in the previous section, it is currently not fully understood how sensitive DNNs are to operational conditions.

\section{Methodology}
\label{sec:methodology}
\begin{figure}[t]
  \centering
  \includegraphics[width=0.5\textwidth]{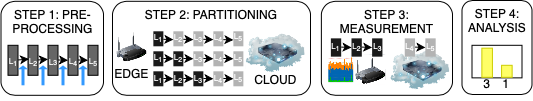}
  \caption{Exhaustive benchmarking method adopted for obtaining data on performance partitioning points of DNNs}
  \label{fig:method}
\end{figure}

The aim of this paper is to carry out exploratory research to address the three questions raised initially in Section~\ref{sec:introduction}.
This requires a methodology for measuring and exhaustively benchmarking the DNN partitions to collect data relevant to the individual layers/blocks of the DNN under varying operational conditions. This section, presents the methodology adopted in this paper and the practical technique used for measurement and benchmarking as shown in Figure~\ref{fig:method}. 

The methodology adopted for obtaining data from the DNNs shown in Table~\ref{tab:pretrainedmodels} has the following four steps:

\textit{Step 1 - Pre-process the DNN and identify suitable partitioning points}:
As discussed in the previous section, a DNN may be a sequential or non-sequential DNN. All layers of a sequential DNN can be partitioned. However, a non-sequential DNN may have parallel paths and therefore there will be fewer suitable partitioning points. In this step, the DNN is pre-processed to identify the valid partitioning points of the DNN. As shown in Table~\ref{tab:pretrainedmodels} non-sequential DNNs have fewer partitioning points than the number of layers; a block of layers will need to be treated as a single entity to avoid partitioning any parallel paths in the DNN. This is to avoid synchronization issues~\cite{couper}.

\textit{Step 2 - Partition the DNN across all suitable partitioning points}:
The DNN is then partitioned across all the above identified partitioning points to ensure that all possible combinations of partitions are available for benchmarking. In this paper, an exhaustive approach is adopted to avoid any assumptions regarding performance as the research reported is an empirical investigation.

\textit{Step 3 - Measure the performance on the edge and cloud resource for varying operational conditions}:
The partitioned DNNs are executed on the edge and cloud resource to measure the end-to-end latencies of discrete individual layers and/or a block of layers of the DNN. For example, consider a sequential DNN with five layers. Then the DNN will be benchmarked for the following 4 combination of partitions: Layer 1 on the edge and Layer 2-5 on the cloud, Layer 1-2 on the edge and Layer 3-5 on the cloud, Layer 1-3 on the edge and Layer 4-5 on the cloud, and Layer 1-4 on the edge and Layer 5 on the cloud. The most performance efficient partition will have the lowest end-to-end latency (which is the sum of the compute time of the partitions on the edge and cloud and the communication time of data from the edge to the cloud). The average of 10 executions of the combination of partitions is noted. 

The operational conditions considered are: (i) CPU stress on the edge: five different stress levels are considered - 0\%, 22\%, 45\%, 67\%, and 90\% CPU utilization; (ii) Memory stress on the edge: five different stress levels are considered - 0\%, 22\%, 45\%, 67\%, and 90\% memory utilization; (iii) Network data transfer rates: 10 Mb/s, 25Mb/s, 37.5Mb/s, and 50Mb/s. For CPU and memory stress on the edge, explicit stress is created on the resource using the Linux tool, namely \texttt{stress}, and the network data transfer rate is controlled using the Linux traffic control tool, namely \texttt{tc}. 
The network data transfer rates are based on different speeds observed in the wide-area network - 50Mb/s is a fast connection available to small businesses in the UK, 25Mb/s is equivalent to the average UK household data download speed, and 10Mb/s is equivalent to the speeds of a more busy network. 

The measurement step is time consuming in that a 1000 executions need to be considered ($5 \times 5 \times 4 \times 10$). One execution of the different DNNs required between 1-7 minutes depending on the depth of the neural network. Therefore, between 1-5 whole days were required for recording the measurements of individual DNNs on each experimental platform considered in the next section.   

\textit{Step 4 - Analyze recorded measurements to identify performance efficient partitions for different scenarios}:  
The measurements obtained from benchmarking are used to determine the most performance efficient partitions. The data is aggregated across multiple dimensions and are presented in the next section.

\section{Experimental Studies}
\label{sec:experimentalstudies}
In this section, the experimental platform used for pursuing studies to identify scenarios when a DNN is sensitive to adaptivity and the results obtained from the study is presented. 

\subsection{Experimental Platform}
The experiments were carried out on four different platforms comprising a cloud and edge processor as shown in Table~\ref{tab:experimentalplatform}.

\begin{table*}[]
\centering
\caption{The four experimental platforms comprising a cloud and edge processor employed in the research}
\label{tab:experimentalplatform}
\begin{tabular}{|c|l|l|c|l|c|l|c|l|c|}
\hline
            & \multicolumn{1}{c|}{} & \multicolumn{4}{c|}{\textbf{Cloud}}  & \multicolumn{4}{c|}{\textbf{Edge}}  \\ \cline{3-10} 
\multirow{-2}{*}{\textbf{Platform}} &
  \multicolumn{1}{c|}{\multirow{-2}{*}{\textbf{OS}}} &
  \multicolumn{1}{c|}{\cellcolor[HTML]{EFEFEF}Processor} &
  \cellcolor[HTML]{EFEFEF}Cores &
  \multicolumn{1}{c|}{\cellcolor[HTML]{EFEFEF}Frequency} &
  \cellcolor[HTML]{EFEFEF}Memory &
  \multicolumn{1}{c|}{\cellcolor[HTML]{EFEFEF}Processor} &
  \cellcolor[HTML]{EFEFEF}Cores &
  \multicolumn{1}{c|}{\cellcolor[HTML]{EFEFEF}Frequency} &
  \cellcolor[HTML]{EFEFEF}Memory \\ \hline
\textbf{$P_1$} & Ubuntu 18.02 LTS      & Intel Xeon E5  & 4 & 2.3 GHz & 8 GB  & ARM Cortex-A72 & 1 & 2.3 GHz & 2 GB \\ \hline
\textbf{$P_2$} & Ubuntu 18.02 LTS      & ARM Cortex-A72 & 8 & 2.3 GHz & 16 GB & ARM Cortex-A72 & 1 & 2.3 GHz & 2 GB \\ \hline
\textbf{$P_3$} & Ubuntu 18.02 LTS      & ARM Cortex-A72 & 8 & 2.3 GHz & 16 GB & Intel Xeon E5  & 2 & 2.3 GHz & 2 GB \\ \hline
\textbf{$P_4$} & Ubuntu 18.02 LTS      & Intel Xeon E5  & 4 & 2.3 GHz & 8 GB  & Intel Xeon E5  & 2 & 2.3 GHz & 2 GB \\ \hline
\end{tabular}
\end{table*}

The benchmarking methodology is implemented in Python and requires Tensorflow 1.5+. Tensorflow~\cite{tensorflow} is used to execute the pre-trained DNNs shown in Table~\ref{tab:pretrainedmodels} that is provided by the Keras\footnote{\url{https://keras.io}}, an open source neural network library. NumPy is used for processing multi-dimensional arrays that are produced as outputs of individual layers or a block of layers. 

The different levels of: (i) CPU stress considered are 0\%, 22\%, 45\%, 67\%, and 90\%, (ii) memory stress considered are 0\%, 22\%, 45\%, 67\%, and 90\%, and (iii) network data transfer rates considered are 10Mb/s, 25Mb/s, 37.5Mb/s and 50Mb/s. However, some of the experiments considered will only present results for fewer stress values and data transfer rates for a meaningful representation of the results. 

\subsection{Results}

Performance efficiency considered in the results is measured as the lowest end-to-end latency when the DNN partitions are executed across the cloud and the edge for a single test image of approximately 150KB size. The values reported for end-to-end latency is the average of ten experimental runs. 

Experimental results are presented to highlight that: 

(i) DNNs are sensitive to operational conditions (to address Q1 posed in Section~\ref{sec:introduction}). For this the percentage of performance efficient partitioning points for DNNs across all combinations of CPU stress, memory stress and network data transfer rates is presented. It is noted that all eight DNNs considered have scope for adaptivity across all potential combinations of CPU stress, memory stress and data transfer rates.

(ii) A performance gain is observed when repartitioning under different operational conditions (to address Q2 presented in Section~\ref{sec:introduction}). This is explored in the context of individual operational conditions. It is noted that although there are performance gains in a number of cases, the overheads in repartitioning and deployment may offset the gain and is likely to depend on the input stream (not considered in this paper).  

(iii) Both individual and a combination of operational conditions affect DNNs (to address Q3 presented in Section~\ref{sec:introduction}). The network conditions affect the DNN performance more directly than CPU and memory stress individually (network conditions have been noted as an important consideration for performance efficiency in connected and autonomous vehicles given that the vehicles move~\cite{cav-01}). There is more impact on DNN performance when a combination of operational conditions are considered. 

(iv) DNN partitioning is sensitive to the hardware architecture employed on the cloud and the edge (to address Q4 posted in Section~\ref{sec:introduction}). The effect of all operational conditions and individual operational conditions on each of the experimental platforms will be considered. It is noted the levels of adaptivity observed vary depending on the network and experimental platform. In general, this highlights the importance of taking the experimental platform into account for partitioning in the real-world, although a specific correlation between the hardware employed on the cloud/edge and its influence on the partitioning point is not noted.

\subsubsection{General Observation on DNN Sensitivity to Operational Conditions}

\begin{figure}[t]
	\centering
	\includegraphics[width=0.115\textwidth]{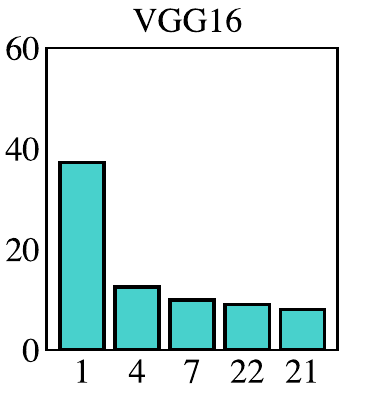}
	\includegraphics[width=0.115\textwidth]{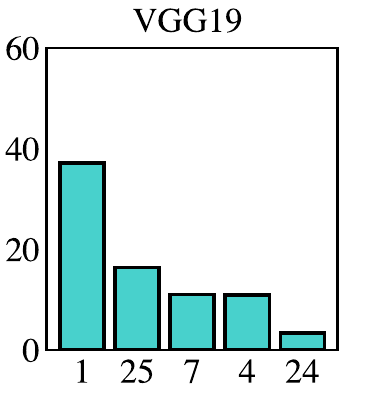}
	\includegraphics[width=0.115\textwidth]{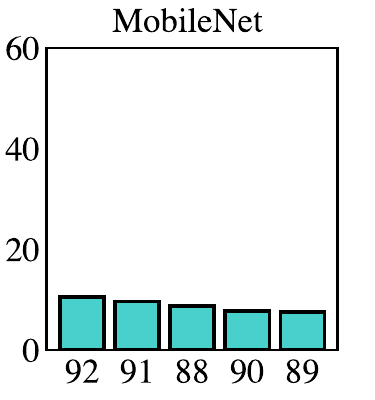}
	\includegraphics[width=0.115\textwidth]{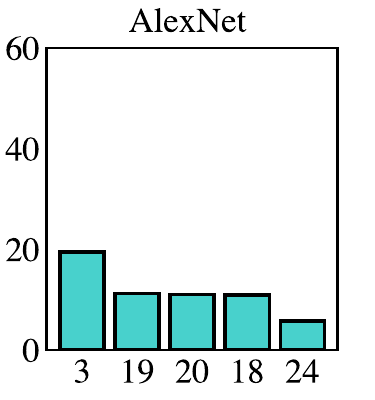}
	\includegraphics[width=0.115\textwidth]{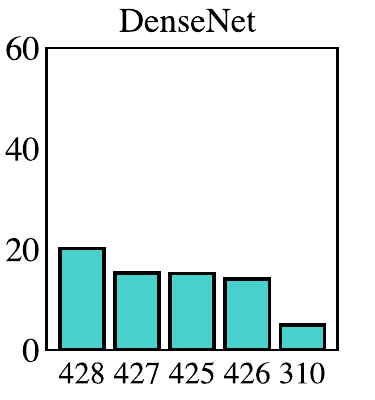}
	\includegraphics[width=0.115\textwidth]{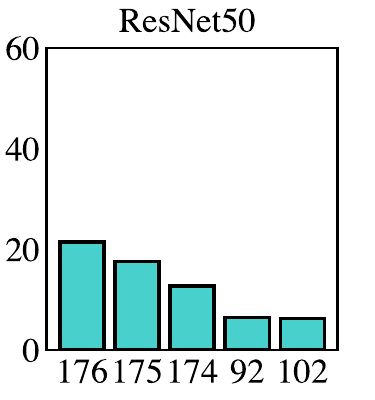}
	\includegraphics[width=0.115\textwidth]{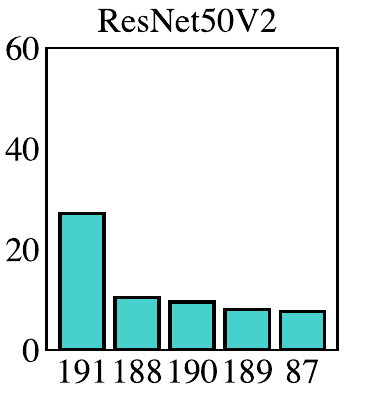}
	\includegraphics[width=0.115\textwidth]{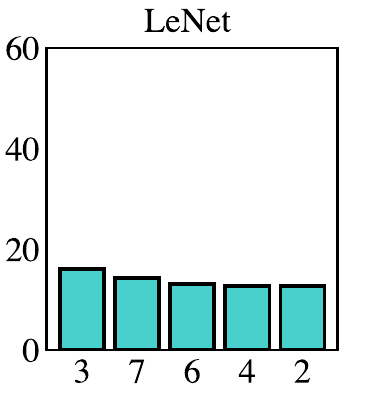}
	\caption{Percentage of performance efficient partitioning points for DNNs across all combinations of CPU stress, memory stress and data transfer rate on all experimental platforms $P_1$, $P_2$, $P_3$ and $P_4$. X-axis shows the partitioning point $n$, which is the $n^{th}$ layer of the DNN where the DNN is partitioned.}
	\label{fig:overalldnns}
\end{figure}

Figure~\ref{fig:overalldnns} shows the percentage of the top 5 performance efficient partitioning points for all combinations of operational conditions considered -- CPU stress, memory stress and network data transfer rates on all experimental platforms $P_1$, $P_2$, $P_3$ and $P_4$. The x-axis shows the partitioning point (the layer after which the DNN is partitioned; for example, a number 81 corresponds to the first partition having layers 1-81 on the edge, and the remaining layers from Layer 82 will be a second partition that is executed on the cloud.

Consider the partition points for VGG16. Regardless of the operational conditions three partitioning points (Layer 1, Layer 4 and Layer 7) result in over 95\% of the most performance efficient partitions. It can then be inferred that in most cases for VGG16, it is highly probable that there will be three options for partitioning. For instance, if the current partition is at Layer 1, then either that is still the most optimal partition, or performance can be optimized by repartitioning with Layer 4 or Layer 7 as the partitioning point. 

Similarly for VGG19 there are three main partitioning points, namely Layer 1, Layer 25 and Layer 7. However, nearly 10\% of operational conditions have Layer 4 as the optimal partition point. 

If ResNet50 is taken as an example, then the partitioning points that result in performance efficient partitions are nearly 50\% of the time at Layer 176 or Layer 175. However, for another 20\% of the cases the partitioning points are at Layer 174 and Layer 92. This is just one evidence of the variation in the partitioning points if the DNN adapts to the changing operational condition. 

Although there is a dominant partitioning point for DenseNet and AlexNet, there are a number of cases in which other partitioning points are optimal. 


\subsubsection{General Observation on DNN Adaptivity and Hardware Architectures}
\begin{figure}[t]
\centering
\includegraphics[width=0.115\textwidth]{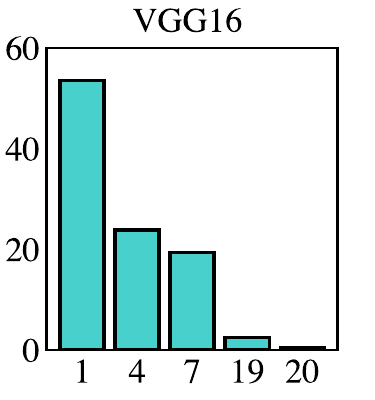}
\includegraphics[width=0.115\textwidth]{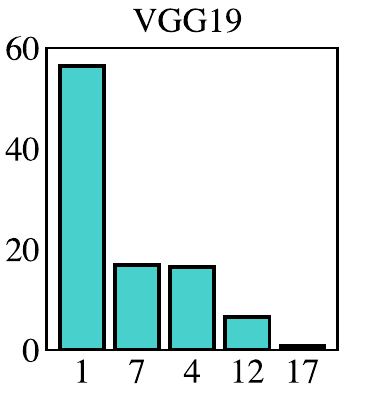}
\includegraphics[width=0.115\textwidth]{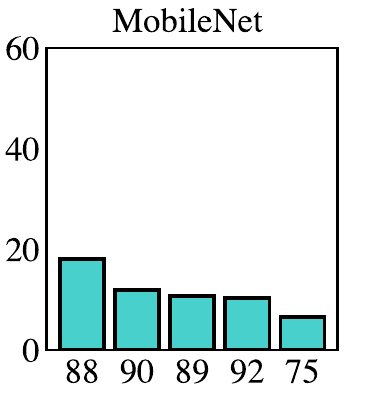}
\includegraphics[width=0.115\textwidth]{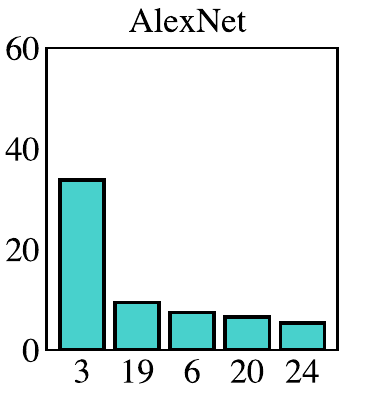}
\includegraphics[width=0.115\textwidth]{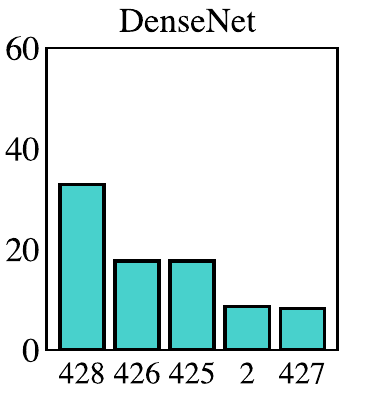}
\includegraphics[width=0.115\textwidth]{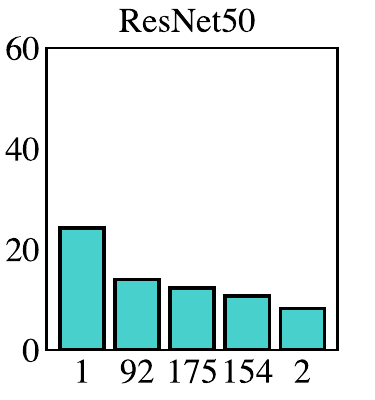}
\includegraphics[width=0.115\textwidth]{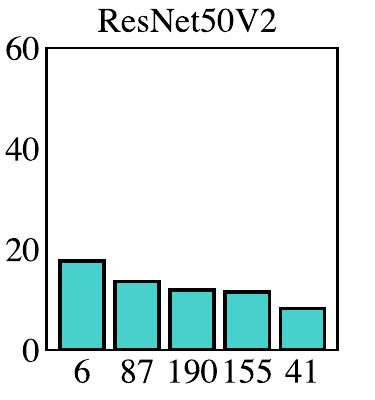}
\includegraphics[width=0.115\textwidth]{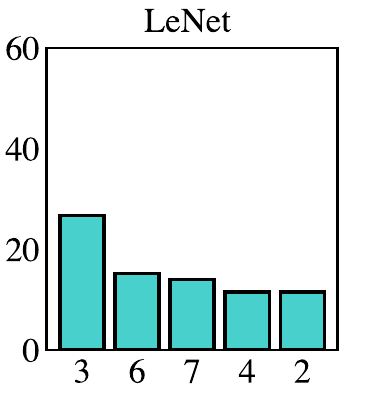}
\caption{Percentage of performance efficient partitioning points for DNNs across all combinations of CPU stress, memory stress and data transfer rate on experimental platform $P_1$. X-axis shows the partitioning point $n$, which is the $n^{th}$ layer of the DNN where the DNN is partitioned.}
\label{fig:overalldnns-p1}
\end{figure}

\begin{figure}[t]
\centering
\includegraphics[width=0.115\textwidth]{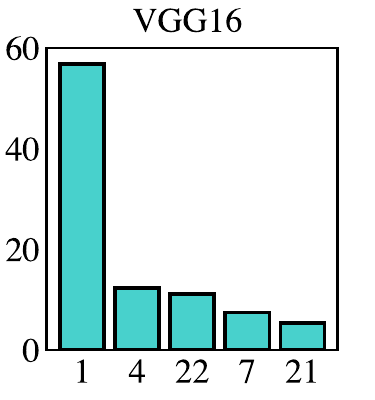}
\includegraphics[width=0.115\textwidth]{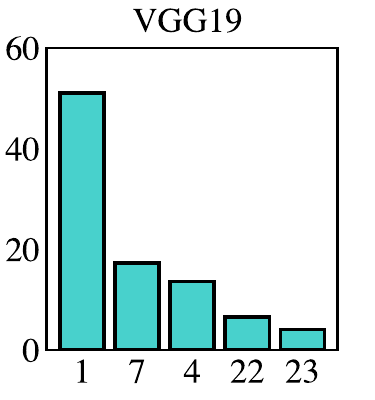}
\includegraphics[width=0.115\textwidth]{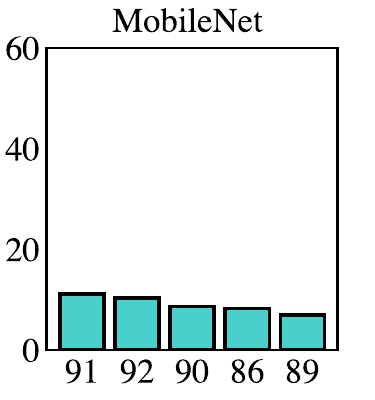}
\includegraphics[width=0.115\textwidth]{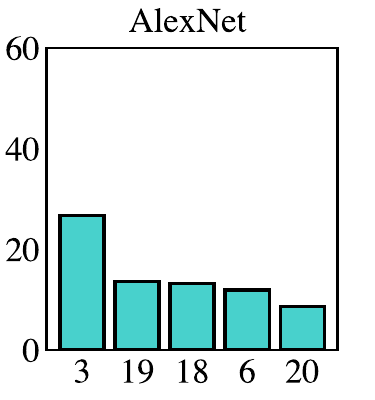}
\includegraphics[width=0.115\textwidth]{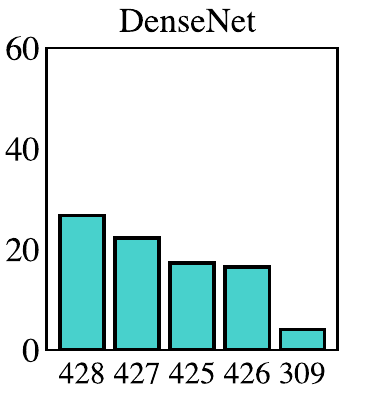}
\includegraphics[width=0.115\textwidth]{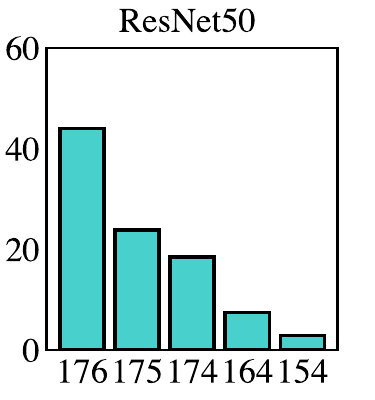}
\includegraphics[width=0.115\textwidth]{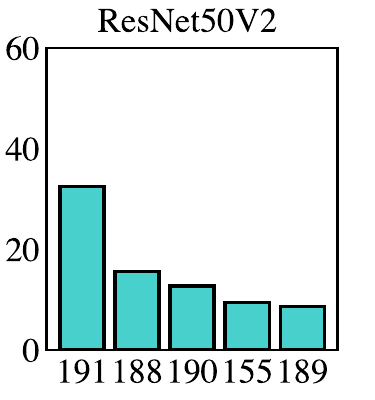}
\includegraphics[width=0.115\textwidth]{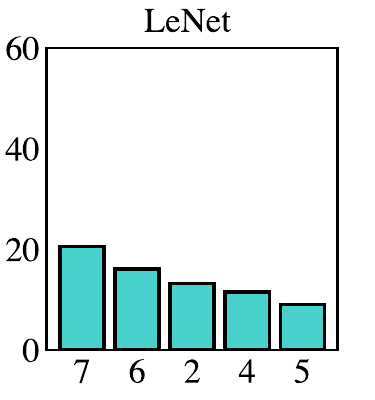}
\caption{Percentage of performance efficient partitioning points for DNNs across all combinations of CPU stress, memory stress and data transfer rate on experimental platform $P_2$. X-axis shows the partitioning point $n$, which is the $n^{th}$ layer of the DNN where the DNN is partitioned.}
\label{fig:overalldnns-p2}
\end{figure}

\begin{figure}[t]
\centering
\includegraphics[width=0.115\textwidth]{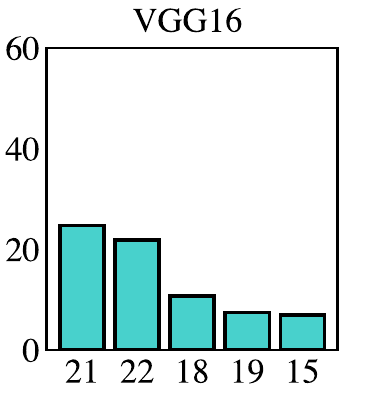}
\includegraphics[width=0.115\textwidth]{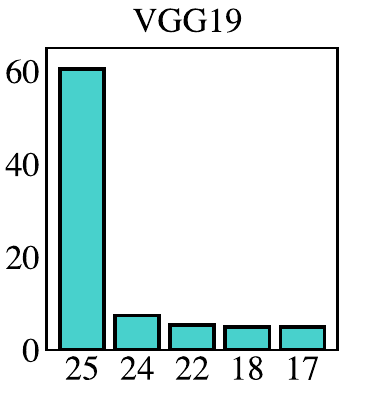}
\includegraphics[width=0.115\textwidth]{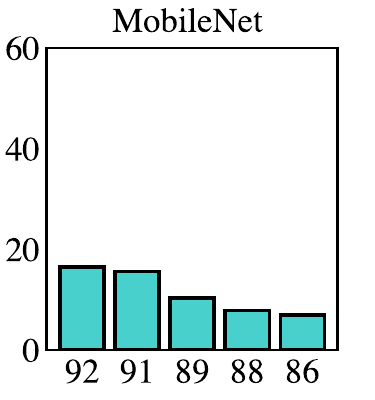}
\includegraphics[width=0.115\textwidth]{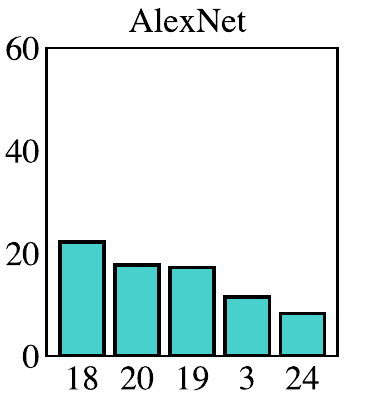}
\includegraphics[width=0.115\textwidth]{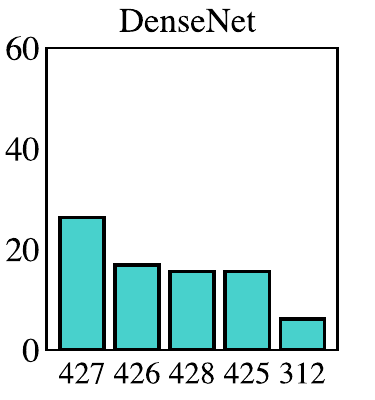}
\includegraphics[width=0.115\textwidth]{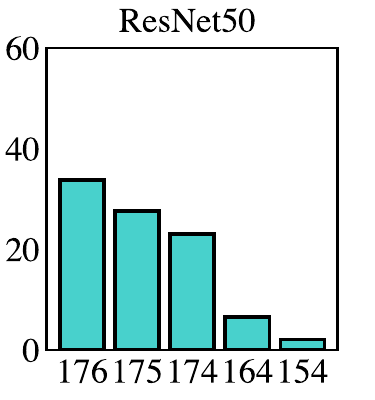}
\includegraphics[width=0.115\textwidth]{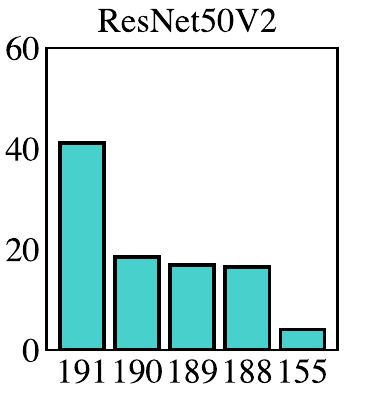}
\includegraphics[width=0.115\textwidth]{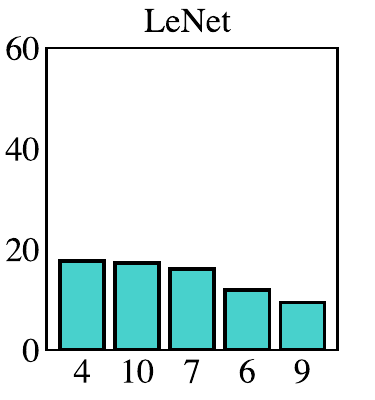}
\caption{Percentage of performance efficient partitioning points for DNNs across all combinations of CPU stress, memory stress and data transfer rate on experimental platform $P_3$. X-axis shows the partitioning point $n$, which is the $n^{th}$ layer of the DNN where the DNN is partitioned.}
\label{fig:overalldnns-p3}
\end{figure}

\begin{figure}[t]
\centering
\includegraphics[width=0.115\textwidth]{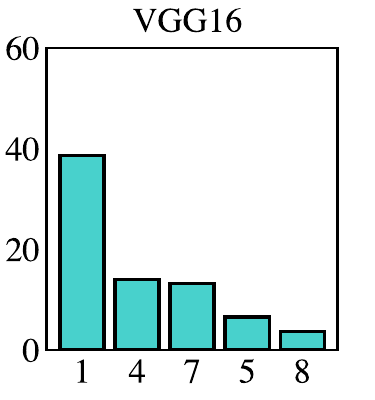}
\includegraphics[width=0.115\textwidth]{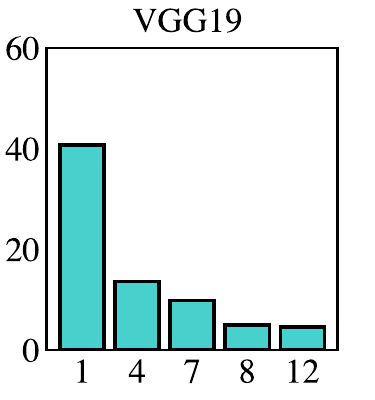}
\includegraphics[width=0.115\textwidth]{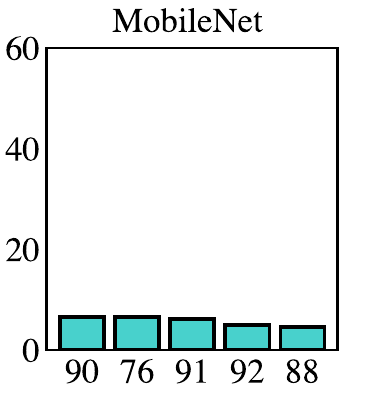}
\includegraphics[width=0.115\textwidth]{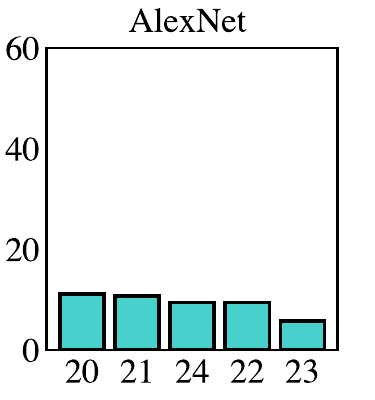}
\includegraphics[width=0.115\textwidth]{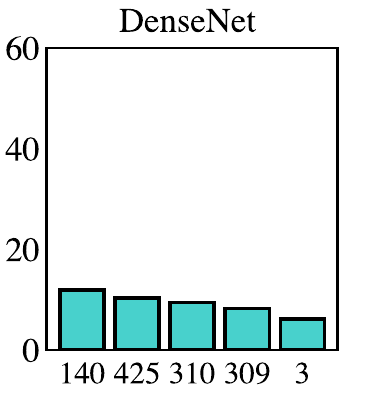}
\includegraphics[width=0.115\textwidth]{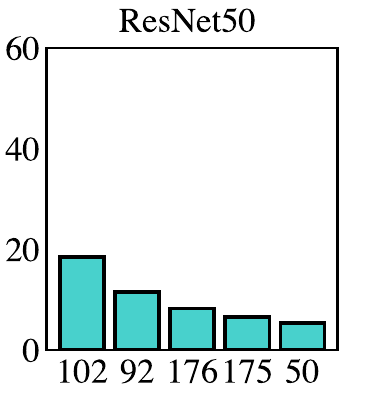}
\includegraphics[width=0.115\textwidth]{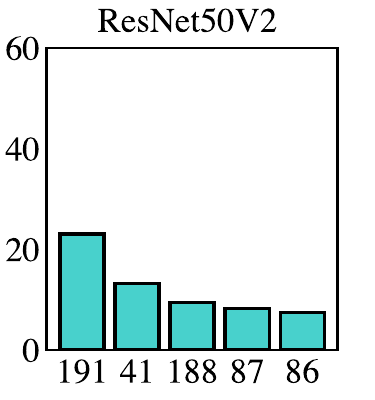}
\includegraphics[width=0.115\textwidth]{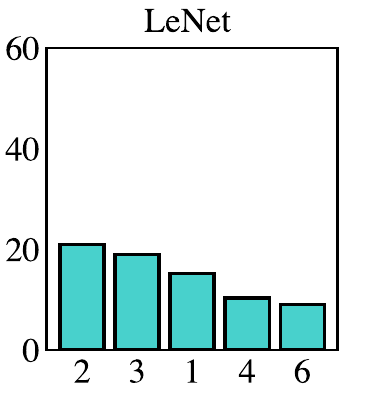}
\caption{Percentage of performance efficient partitioning points for DNNs across all combinations of CPU stress, memory stress and data transfer rate on experimental platform $P_4$. X-axis shows the partitioning point $n$, which is the $n^{th}$ layer of the DNN where the DNN is partitioned.}
\label{fig:overalldnns-p4}
\end{figure}

Figure~\ref{fig:overalldnns-p1}, Figure~\ref{fig:overalldnns-p2}, Figure~\ref{fig:overalldnns-p3} and Figure~\ref{fig:overalldnns-p4} shows the percentage of the top 5 performance efficient partitioning points for all combinations of operational conditions considered -- CPU stress, memory stress and network data transfer rates on the experimental platforms $P_1$, $P_2$, $P_3$ and $P_4$ respectively.

To demonstrate the sensitivity to hardware architectures on the edge, we compare the results from platforms $P_1$ and $P_4$ (same Intel processor on the cloud, but different edge processors; Figure~\ref{fig:overalldnns-p1} and Figure~\ref{fig:overalldnns-p4}) and from $P_2$ and $P_3$ (same ARM processor on the cloud, but different edge processors; Figure~\ref{fig:overalldnns-p2} and Figure~\ref{fig:overalldnns-p3}). 

Consider AlexNet and ResNet50 on $P_1$ and $P_4$. For both DNNs the dominant partitioning points on the Arm edge processor are early in the network (Layer 3 for AlexNet and Layer 1 for ResNet50), but on the Intel edge processor the dominant partitioning points are further down in the network. Similarly, consider VGG16 and VGG19 on $P_2$ and $P_3$. For both DNNs the dominant partitioning points on the Arm edge processor are earlier in the network, where as on the Intel edge processor appears further down in the network. 

The results suggest that there is a limited influence of the cloud processor on the partitioning point at the edge. 

\subsubsection{DNN Adaptivity and Individual Operational Conditions}
The effect of CPU stress, memory stress and network data transfer rate individually on DNN adaptivity is considered. The results presented for varying individual operational condition do not additionally stress other conditions. For example, the experiments for identifying the sensitivity of CPU stress to DNN adaptivity has no additional memory stress applied and has the maximum network data transfer rate.

\begin{figure}[t]
\centering
\includegraphics[width=0.115\textwidth]{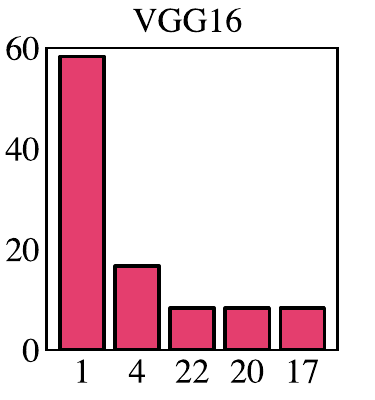}
\includegraphics[width=0.115\textwidth]{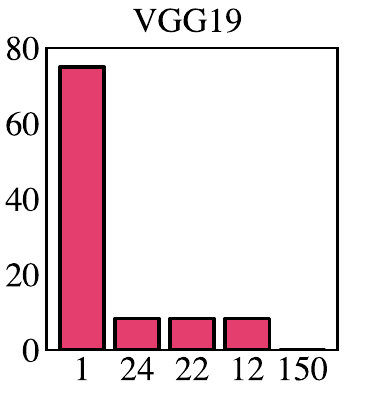}
\includegraphics[width=0.115\textwidth]{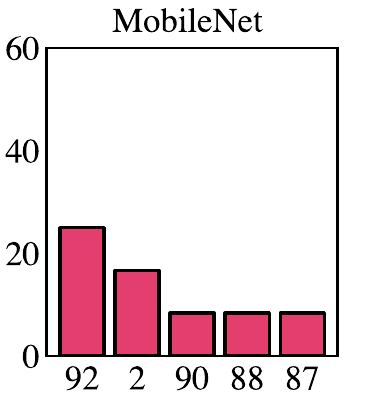}
\includegraphics[width=0.115\textwidth]{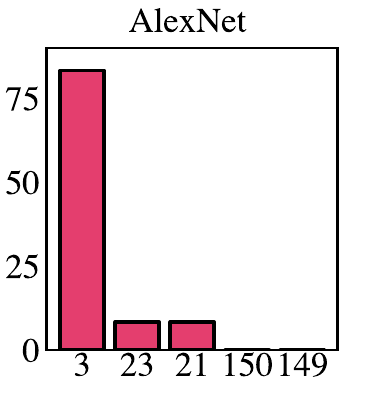}
\includegraphics[width=0.115\textwidth]{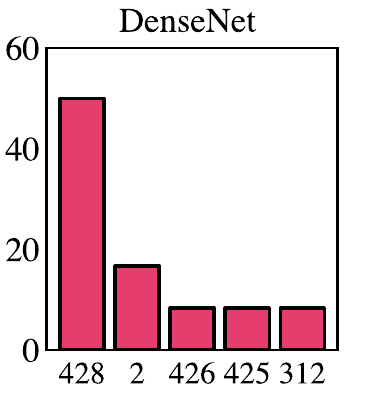}
\includegraphics[width=0.115\textwidth]{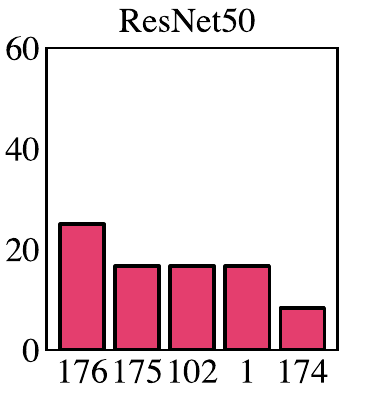}
\includegraphics[width=0.115\textwidth]{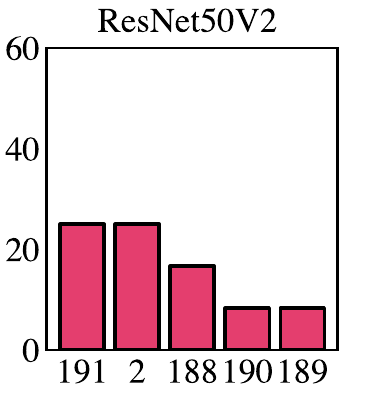}
\includegraphics[width=0.115\textwidth]{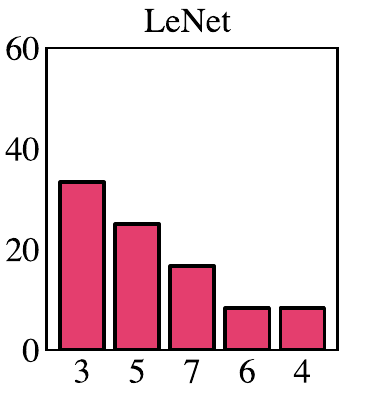}
\caption{Percentage of performance efficient partitioning points for the DNNs for different CPU stress when no additional memory stress is applied on the edge and the network data transfer rate is 50Mb/s on all experimental platforms $P_1$, $P_2$, $P_3$ and $P_4$. X-axis shows the partitioning point $n$, which is the $n^{th}$ layer of the DNN where the DNN is partitioned.}
\label{fig:CPUstress}
\end{figure}

\begin{figure}[t]
\centering
\includegraphics[width=0.115\textwidth]{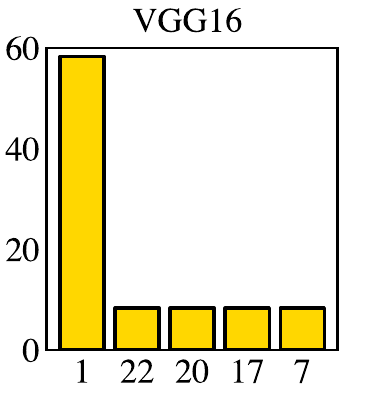}
\includegraphics[width=0.115\textwidth]{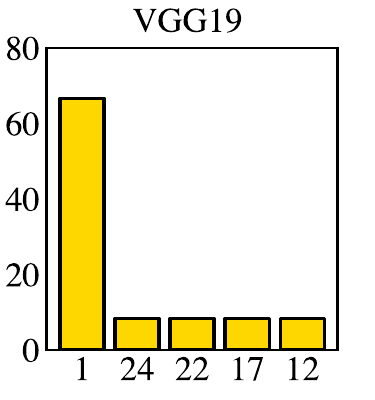}
\includegraphics[width=0.115\textwidth]{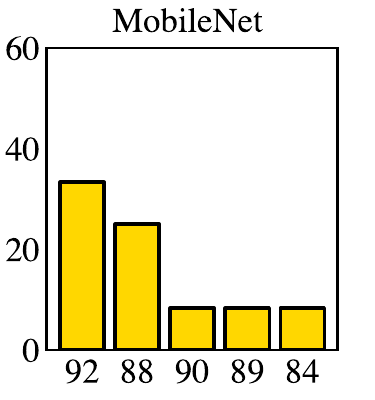}
\includegraphics[width=0.115\textwidth]{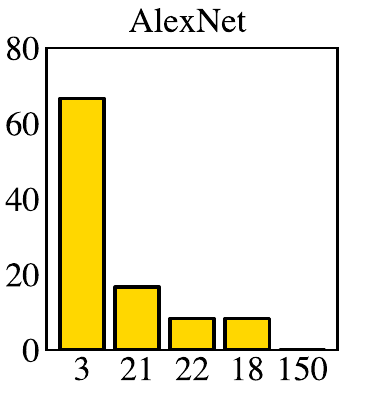}
\includegraphics[width=0.115\textwidth]{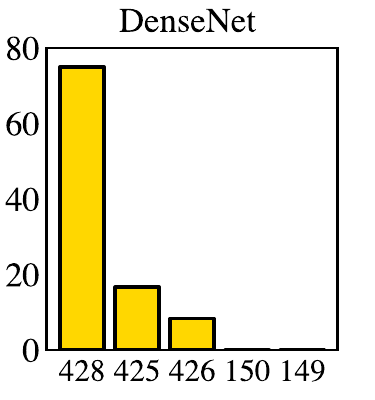}
\includegraphics[width=0.115\textwidth]{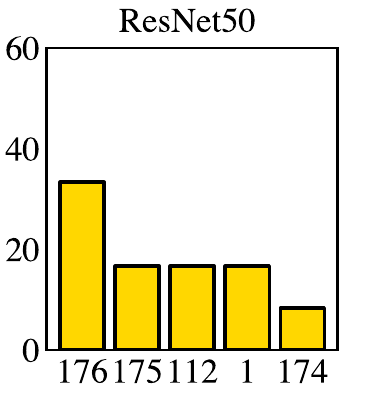}
\includegraphics[width=0.115\textwidth]{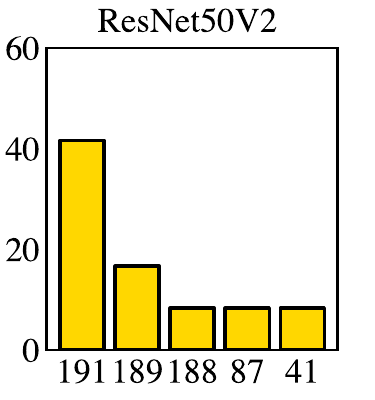}
\includegraphics[width=0.115\textwidth]{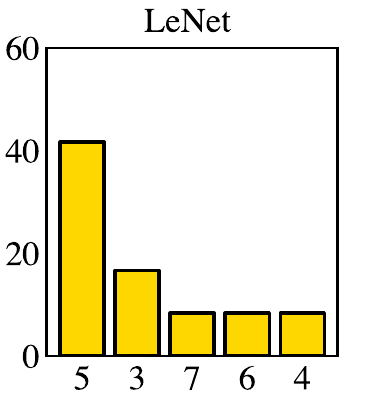}
\caption{Percentage of performance efficient partitioning points for the DNNs for different memory stress when no additional CPU stress is applied on the edge and the network data transfer rate is 50Mb/s on all experimental platforms $P_1$, $P_2$, $P_3$ and $P_4$. X-axis shows the partitioning point $n$, which is the $n^{th}$ layer of the DNN where the DNN is partitioned.}
\label{fig:memorystress}
\end{figure}

\begin{figure}[t]
\centering
\includegraphics[width=0.115\textwidth]{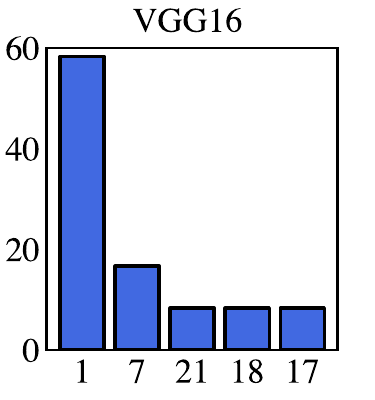}
\includegraphics[width=0.115\textwidth]{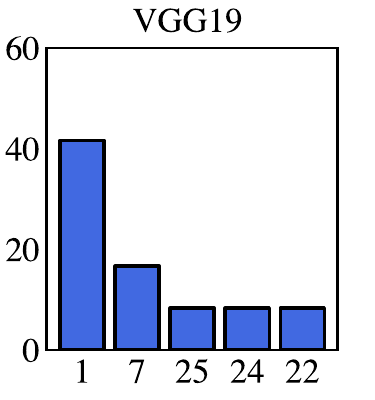}
\includegraphics[width=0.115\textwidth]{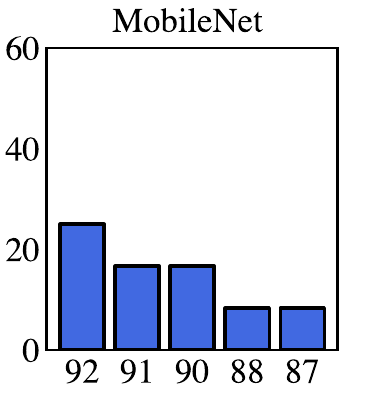}
\includegraphics[width=0.115\textwidth]{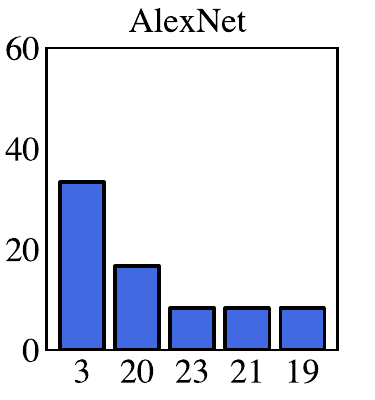}
\includegraphics[width=0.115\textwidth]{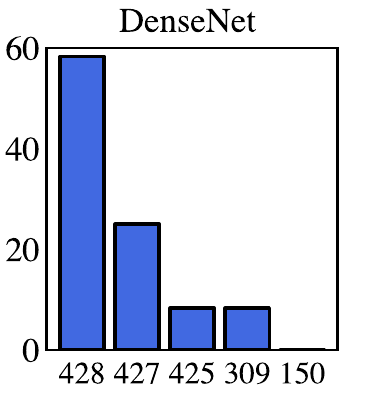}
\includegraphics[width=0.115\textwidth]{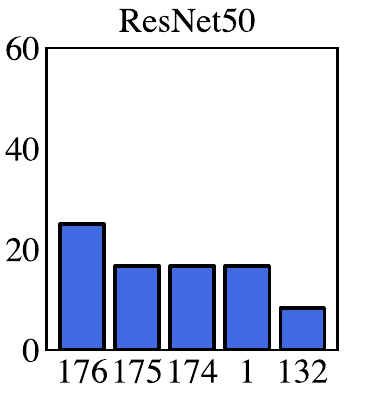}
\includegraphics[width=0.115\textwidth]{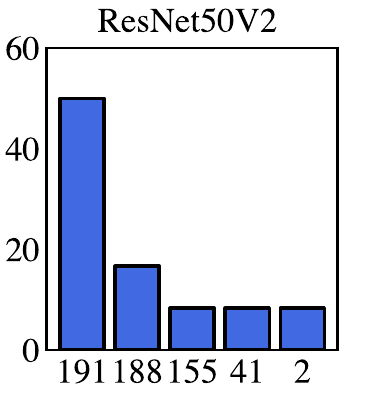}
\includegraphics[width=0.115\textwidth]{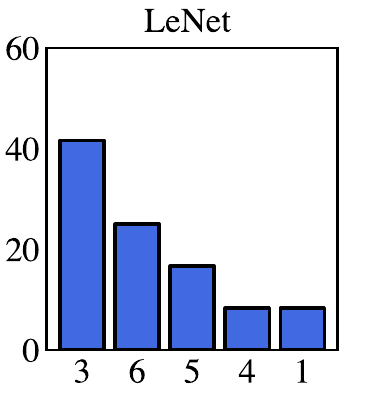}
\caption{Percentage of performance efficient partitioning points for the DNNs for different network data transfer rates when no additional CPU or memory stress is applied on the edge on all experimental platforms $P_1$, $P_2$, $P_3$ and $P_4$. X-axis shows the partitioning point $n$, which is the $n^{th}$ layer of the DNN where the DNN is partitioned.}
\label{fig:networkstress}
\end{figure}

Two comparisons are relevant and are considered here. Firstly, comparing DNN sensitivity to CPU stress, memory stress and network data transfer rates individually against Figure~\ref{fig:overalldnns}. This will highlight the effect of the individual operational conditions on DNN partitioning points against all combinations of the operational conditions. Secondly, comparing the graphs that show the impact of DNN partitioning points on CPU stress, memory stress and network data transfer rates against each other. This will highlight the DNNs that are sensitive to specific individual operational conditions. 

\textit{Sensitivity to CPU stress}:
Figure~\ref{fig:CPUstress} shows the percentage of the top performance efficient partitioning points for different values of CPU stress on the edge, when no additional memory stress is applied and the network data transfer rate is 50Mb/s. 
Comparing with Figure~\ref{fig:overalldnns}, it is noted that AlexNet has only 3 optimal partitioning points with layer 3 being the most frequent and VGG19 has only 4 optimal partitioning points. There are fewer optimal partitioning points for these DNNs for different CPU stress; it was noted that when using the Intel processor on the edge, the number of partitioning points increased for these networks (results not shown since the graphs are exhaustive). The other partitioning points, for example layers 150 and 149 of AlexNet shown in Figure~\ref{fig:CPUstress} are the sub-optimal partitioning points in decreasing order. For VGG19 it is noted that the partitioning point that is most prominent across the search space (Layer 1; Figure~\ref{fig:overalldnns}) is the optimal partition point under maximum CPU stress (most layers have to reside on the cloud) and remains the optimal partitioning point, regardless of the underlying hardware architecture.
DenseNet demonstrates the impact of CPU stress on optimal partitioning points effectively. A number of layers are sensitive to CPU stress, but are less sensitive to memory stress and network data transfer rates. 

\textit{Sensitivity to memory stress}:
Figure~\ref{fig:memorystress} shows the percentage of the top performance efficient partitioning points for different values of memory stress on the edge, when no additional CPU stress is applied and the network data transfer rate is 50Mb/s. 
Similar to the sensitivity to CPU stress, VGG16 and VGG19 show little sensitivity, with layer 1 being overwhelmingly the most frequent optimal partition point. AlexNet has shown broadly the same optimal partitioning points as in Figure~\ref{fig:CPUstress}. VGG16 and ResNet50 has a similar profile as seen in Figure~\ref{fig:CPUstress}. Both CPU stress and memory stress have a similar diversity of partitioning points.  

\textit{Sensitivity to network data transfer rate}:
Figure~\ref{fig:networkstress} shows the percentage of the top performance efficient partitioning points for different network data transfer rates when no additional CPU or memory stress is applied on the edge.
Two layers, Layer 1 and Layer 7 are prominent partitioning points for VGG16. 
In the case of AlexNet, Layer 23 appears as an optimal partitioning point although it does not appear as a top five across all combinations of the operational conditions. 
The optimal partitioning point Layer 428 in Figure~\ref{fig:overalldnns} becomes even more prominent for DenseNet in Figure~\ref{fig:networkstress}.


\begin{table*}[]
\centering
\caption{Sensitivity of DNNs to individual operational conditions on experimental platforms, $P_1$, $P_2$, $P_3$ and $P_4$; Y - Yes, N - No}
\label{tab:sensitivity-compare-individual}
\begin{tabular}{|l|
>{\columncolor[HTML]{EFEFEF}}c |
>{\columncolor[HTML]{EFEFEF}}c |
>{\columncolor[HTML]{EFEFEF}}c |
>{\columncolor[HTML]{EFEFEF}}c |c|c|c|c|
>{\columncolor[HTML]{EFEFEF}}c |
>{\columncolor[HTML]{EFEFEF}}c |
>{\columncolor[HTML]{EFEFEF}}c |
>{\columncolor[HTML]{EFEFEF}}c |}
\hline
\multicolumn{1}{|c|}{} &
  \multicolumn{4}{c|}{\cellcolor[HTML]{EFEFEF}\textbf{CPU Stress}} &
  \multicolumn{4}{c|}{\cellcolor[HTML]{EFEFEF}\textbf{Memory Stress}} &
  \multicolumn{4}{c|}{\cellcolor[HTML]{EFEFEF}\textbf{Network Data Transfer}} \\ \cline{2-13} 
\multicolumn{1}{|c|}{\multirow{-2}{*}{\textbf{DNN Model}}} &
  \textbf{$P_1$} &
  \textbf{$P_2$} &
  \textbf{$P_3$} &
  \textbf{$P_4$} &
  \textbf{$P_1$} &
  \textbf{$P_2$} &
  \textbf{$P_3$} &
  \textbf{$P_4$} &
  \textbf{$P_1$} &
  \textbf{$P_2$} &
  \textbf{$P_3$} &
  \textbf{$P_4$} \\ \hline
ResNet50V2 & N & Y & Y & Y & Y & Y & N & Y & Y & Y & Y & Y \\ \hline
MobileNet  & Y & Y & Y & Y & N & N & Y & Y & Y & Y & Y & Y \\ \hline
ResNet50   & Y & Y & Y & Y & Y & N & Y & Y & Y & N & Y & Y \\ \hline
VGG16      & Y & N & Y & N & Y & N & Y & Y & Y & N & Y & Y \\ \hline
VGG19      & Y & N & Y & Y & Y & N & Y & Y & Y & N & Y & Y \\ \hline
DenseNet   & Y & Y & Y & Y & N & Y & Y & Y & N & Y & Y & Y \\ \hline
AlexNet    & N & N & N & Y & N & N & Y & Y & Y & Y & Y & Y \\ \hline
LeNet      & N & Y & Y & Y & Y & Y & Y & Y & N & Y & Y & Y \\ \hline
\end{tabular}
\end{table*}

The comparison of the DNNs among the three individual operational conditions on all experimental platforms is summarized in Table~\ref{tab:sensitivity-compare-individual}.
When CPU stress is considered MobileNet, ResNet50, ResNet50V2 and DenseNet are adaptive on all experimental platforms. 
Similarly, for memory stress MobileNet, ResNet50, ResNet50V2 and LeNet are adaptive to all experimental platforms and for network data transfer ResNet50V2, ResNet50 and MobileNet are adaptive to all experimental platforms. 
All networks show sensitivity to individual operational conditions across all experimental platforms, however the VGG networks usually perform best when layer 1 is chosen as the partition point.
The choice of processor on the edge affects the optimal partitioning points for all the networks, with the optimal partitioning point moving deeper into the network when an Intel architecture is used. This highlights that the underlying hardware influences the partitioning point in response to individual operational conditions. However, specific patterns of the sensitivity to hardware architectures cannot be obtained. 


\begin{table*}[]
\centering
\caption{Effect of CPU stress on end-to-end latency (seconds) on experimental platform $P_1$; partitioning layer is shown in brackets beside the end-to-end latency values. End-to-end latency is shown for 0\%, 45\% and 90\% CPU stress when partitioning layer is the same as for CPU stress is 0\%. End-to-end latency for the best DNN partition when CPU stress is 45\% and 90\% along with the partitioning layer is shown. The performance gain of best partition is shown. AlexNet, ResNet50V2 and LeNet are not included as they are not sensitive to CPU stress (refer to Table~\ref{tab:sensitivity-compare-individual}).}
\label{tab:cpu-stress-fullisolation}
\begin{tabular}{|l|
>{\columncolor[HTML]{EFEFEF}}l |
>{\columncolor[HTML]{FFFFEC}}l |
>{\columncolor[HTML]{FFFFEC}}l |
>{\columncolor[HTML]{FFFFEC}}c |
>{\columncolor[HTML]{FFF4E6}}l |
>{\columncolor[HTML]{FFF4E6}}l |
>{\columncolor[HTML]{FFF4E6}}c |}
\hline
\multicolumn{1}{|c|}{\textbf{DNN Model}} &
  \multicolumn{1}{c|}{\cellcolor[HTML]{EFEFEF}\textbf{\begin{tabular}[c]{@{}c@{}}0\%\\ (best partition)\end{tabular}}} &
  \multicolumn{1}{c|}{\cellcolor[HTML]{FFFFEC}\textbf{45\%}} &
  \multicolumn{1}{c|}{\cellcolor[HTML]{FFFFEC}\textbf{\begin{tabular}[c]{@{}c@{}}45\%\\ (best partition)\end{tabular}}} &
  \textbf{Gain (\%)} &
  \multicolumn{1}{c|}{\cellcolor[HTML]{FFF4E6}\textbf{90\%}} &
  \multicolumn{1}{c|}{\cellcolor[HTML]{FFF4E6}\textbf{\begin{tabular}[c]{@{}c@{}}90\% \\ (best partition)\end{tabular}}} &
  \textbf{Gain (\%)} \\ \hline
VGG16       & 1.048 (1)   & 3.236 (1)   & 2.022 (4)   & 37.51 & 3.260 (1)   & 2.349 (4)  & 27.94     \\ \hline
VGG19       & 1.261 (1)   & 3.395 (1)   & 2.417 (4)   & 28.81 & 3.468 (1)    & 2.564 (1)  & 26.10 \\ \hline
MobileNet   & 0.280 (88)  & 0.524 (88)  & 0.291 (83)  & 44.47  & 0.554 (88)  & 0.370 (78) & 33.21  \\ \hline
DenseNet    & 1.179 (428) & 1.554 (428) & 1.276 (2) & 17.89 & 2.254 (428) & 1.296 (2) & 42.50  \\ \hline
ResNet50    & 0.952 (112) & 1.046 (112) & 0.960 (1)  & 8.22  & 1.827 (112) & 0.953 (5)  & 47.84 \\ \hline
\end{tabular}
\end{table*}

\textit{Performance Gain}:
Table~\ref{tab:cpu-stress-fullisolation} shows the performance gain of repartitioning DNNs for different CPU stress levels on experimental platform $P_1$ (similar results obtained from other platforms are not presented since they are exhaustive). The table shows the end-to-end latency (in seconds) and the partitioning layer for 0\%, 45\% and 90\% CPU stress at the edge. The partitioning layer is the same for 45\% and 90\% as it is for when there is no CPU stress. The general trend is that the end-to-end latency increases with increasing CPU stress. The table then shows the end-to-end latency of the best DNN partition at a different partition layer for 45\% and 90\% CPU stress (shown in the table as best partition). The performance gain of the best partition over using a static partitioning layer (best partitioning point when CPU stress is 0\%) is indicated in the table. 

Consider DenseNet as an example in Table~\ref{tab:cpu-stress-fullisolation}. The performance gain is immediately evident. If the original partition is used (at Layer 428; optimal partition when CPU stress is 0\%), then the end-to-end latency of this distributed DNN would be 2.254 seconds when the CPU stress is 90\%. However, repartitioning at Layer 2, results in a DNN with 1.296 seconds end-to-end latency resulting in a 42.50\% performance gain. A smaller gain of 17.89\% is noted when the DNN is repartitioned at Layer 2 if there is 45\% CPU stress. 
The performance gain is only an indicator that there is benefit to repartitioning and DNNs are sensitive to adaptivity. 

\begin{table*}[]
\centering
\caption{Effect of memory stress on end-to-end latency (seconds) on experimental platform $P_1$; partitioning layer is shown in brackets beside the end-to-end latency values. End-to-end latency is shown for 0\%, 45\% and 90\% memory stress for the partitioning layer when memory stress is 0\%. End-to-end latency for the best DNN partition when memory stress is 45\% and 90\% along with the partitioning layer is shown. The performance gain of the best partition is shown. MobileNet, AlexNet and DenseNet are not included as they are not sensitive to memory stress (refer to Table~\ref{tab:sensitivity-compare-individual}).}
\label{tab:memory-stress-fullisolation}
\begin{tabular}{|l|
>{\columncolor[HTML]{EFEFEF}}l |
>{\columncolor[HTML]{FFFFEC}}l |
>{\columncolor[HTML]{FFFFEC}}l |
>{\columncolor[HTML]{FFFFEC}}c |
>{\columncolor[HTML]{FFF4E6}}l |
>{\columncolor[HTML]{FFF4E6}}l |
>{\columncolor[HTML]{FFF4E6}}c |}
\hline
\multicolumn{1}{|c|}{\textbf{DNN Model}} &
  \multicolumn{1}{c|}{\cellcolor[HTML]{EFEFEF}\textbf{\begin{tabular}[c]{@{}c@{}}0\%\\ (best partition)\end{tabular}}} &
  \multicolumn{1}{c|}{\cellcolor[HTML]{FFFFEC}\textbf{45\%}} &
  \multicolumn{1}{c|}{\cellcolor[HTML]{FFFFEC}\textbf{\begin{tabular}[c]{@{}c@{}}45\%\\ (best partition)\end{tabular}}} &
  \textbf{Gain (\%)} &
  \multicolumn{1}{c|}{\cellcolor[HTML]{FFF4E6}\textbf{90\%}} &
  \multicolumn{1}{c|}{\cellcolor[HTML]{FFF4E6}\textbf{\begin{tabular}[c]{@{}c@{}}90\% \\ (best partition)\end{tabular}}} &
  \textbf{Gain (\%)} \\ \hline
VGG16       & 1.048 (1)   & 1.897 (1)   & 1.817 (4)   & 4.22     & 2.059 (1)   & 2.059 (1)   & 0     \\ \hline
VGG19       & 1.261 (1)   & 3.389 (1)   & 2.040 (4)    & 39.81  & 3.385 (1)   & 2.039 (1)   & 39.76  \\ \hline
ResNet50    & 0.952 (112) & 1.054 (112) & 0.945 (1)  & 10.34   & 1.048 (112) & 0.941 (1)  & 10.31  \\ \hline
ResNet50V2 & 0.678 (2)   & 0.684 (2)   & 0.676 (1)  & 1.17 & 2.094 (2)   & 0.753 (87)  & 64.04 \\ \hline
LeNet       & 0.008 (3)   & 0.010 (3)   & 0.009 (6)   & 10.00 & 0.012 (3)   & 0.011 (7)   & 8.33    \\ \hline
\end{tabular}
\end{table*}

Table~\ref{tab:memory-stress-fullisolation} shows the performance gain of repartitioning DNNs for different memory stress levels. The table shows the end-to-end latency (in seconds) and the partitioning layer for 0\%, 45\% and 90\% memory stress at the edge. The partitioning layer is the same for 45\% and 90\% as it is for when there is no memory stress. The general trend is that the end-to-end latency increases with increasing memory stress. The table then shows the end-to-end latency of the best DNN partition at a different partition layer for 45\% and 90\% memory stress (shown in the table as best partition). The performance gain by using a best partition layer over using a static partitioning layer (best partitioning point when memory stress is 0\%) is indicated in the table. 
A noteworthy gain from repartitioning is observed for VGG19 and ResNet50 V2. The DNNs tend to show less sensitivity to memory stress when compared to CPU stress. 

\begin{table*}[]
\centering
\caption{Effect of network data transfer rates between the edge and the cloud on end-to-end latency (seconds) when using experimental platform $P_1$; partitioning layer is shown in brackets beside the end-to-end latency values. End-to-end latency is shown for 10Mb/s, 25Mb/s and 50Mb/s data transfer rate (partitioning layer is the same as when transfer is 50Mb/s). End-to-end latency for the best DNN partition when network data transfer rate is 25Mb/s and 10Mb/s along with the partitioning layer is shown. The performance gain of the best partition is shown. LeNet and DenseNet are not shown as they are not sensitive to network transfer rates (refer to Table~\ref{tab:sensitivity-compare-individual}).}
\label{tab:network-stress-fullisolation}
\begin{tabular}{|l|
>{\columncolor[HTML]{EFEFEF}}l |
>{\columncolor[HTML]{FFFFEC}}l |
>{\columncolor[HTML]{FFFFEC}}l |
>{\columncolor[HTML]{FFFFEC}}c |
>{\columncolor[HTML]{FFF4E6}}l |
>{\columncolor[HTML]{FFF4E6}}l |
>{\columncolor[HTML]{FFF4E6}}c |}
\hline
\multicolumn{1}{|c|}{\textbf{DNN Model}} &
  \multicolumn{1}{c|}{\cellcolor[HTML]{EFEFEF}\textbf{\begin{tabular}[c]{@{}c@{}}50Mb/s\\ (best partition)\end{tabular}}} &
  \multicolumn{1}{c|}{\cellcolor[HTML]{FFFFEC}\textbf{25Mb/s}} &
  \multicolumn{1}{c|}{\cellcolor[HTML]{FFFFEC}\textbf{\begin{tabular}[c]{@{}c@{}}25Mb/s\\ (best partition)\end{tabular}}} &
  \textbf{Gain (\%)} &
  \multicolumn{1}{c|}{\cellcolor[HTML]{FFF4E6}\textbf{10Mb/s}} &
  \multicolumn{1}{c|}{\cellcolor[HTML]{FFF4E6}\textbf{\begin{tabular}[c]{@{}c@{}}10Mb/s \\ (best partition)\end{tabular}}} &
  \textbf{Gain (\%)} \\ \hline
VGG16       & 1.048 (1)   & 1.131 (1)   & 1.131 (1)   & 0  & 3.606 (1)   & 3.004 (6)   & 16.70 \\ \hline
VGG19       & 1.261 (1)   & 5.585 (1)   & 2.489 (7)   & 55.43    & 14.307 (1)   & 3.254 (7)   & 77.25     \\ \hline
MobileNet   & 0.280 (88)  & 0.319 (88)  & 0.293 (90)  & 8.15 & 0.384 (88)  & 0.346 (90)  & 9.89 \\ \hline
AlexNet     & 0.107 (3)   & 0.446 (3)   & 0.278 (23)   & 37.67     & 0.516 (3)   & 0.345 (10)   & 33.14 \\ \hline
ResNet50    & 0.952 (112) & 1.174 (112) & 0.971 (1)   & 17.29  & 1.632 (112) & 1.196 (1)  & 26.72 \\ \hline
ResNet50V2 & 0.678 (2)   & 0.721 (2)   & 0.717 (1)  & 0.55  & 2.295 (2)    & 0.974 (155) & 57.56  \\ \hline
\end{tabular}
\end{table*}

Table~\ref{tab:network-stress-fullisolation} shows the performance gain of repartitioning DNNs for different network data transfer rates. The table shows the end-to-end latency (in seconds) and the partitioning layer for 50Mb/s, 25Mb/s, and 10Mb/s between the edge and the cloud. The partitioning layer is the same for 25Mb/s and 10Mb/s as it is when there is maximum available bandwidth. The general trend is that the end-to-end latency increases with decreasing data transfer rates. The table then shows the end-to-end latency of the best DNN partition at a different partition layer for 25Mb/s and 10Mb/s (shown in the table as best partition). The performance gain of employing the best partition over using a static partitioning layer (best partitioning point when network data transfer rate is 50Mb/s) is highlighted in the table. 

It is immediately inferred that the performance gain for the selected DNNs in response to different network data transfer rates is greater than for CPU or memory stress. 
AlexNet, that was not sensitive to CPU and memory stress, is more sensitive to changing network conditions and benefits from repartitioning; up to 37.67\% gains are noted. 
Although VGG16 is sensitive to network conditions, there are instances when there is no performance gain. This is because best partition does not provide any added benefit in this particular case.

\subsubsection{DNN Adaptivity and a Combination of Operational Conditions}

\begin{figure}[t]
\centering
\includegraphics[width=0.115\textwidth]{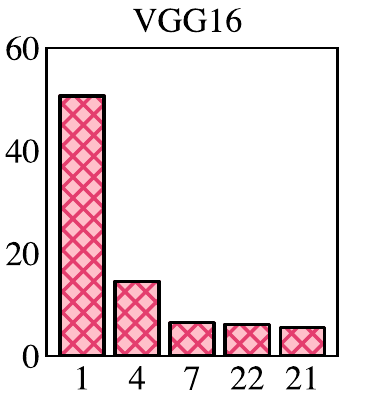}
\includegraphics[width=0.115\textwidth]{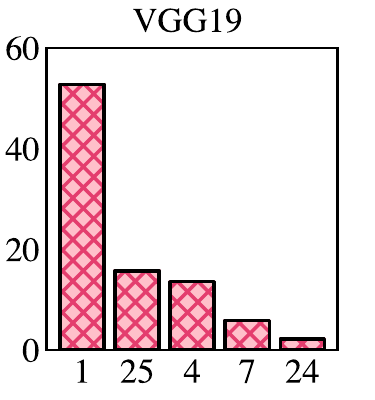}
\includegraphics[width=0.115\textwidth]{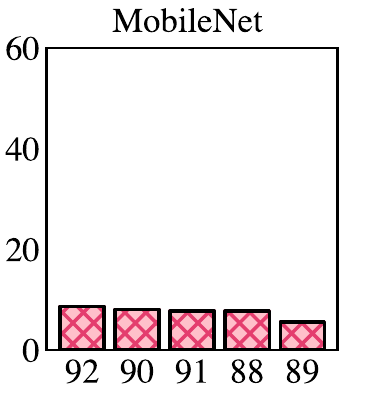}
\includegraphics[width=0.115\textwidth]{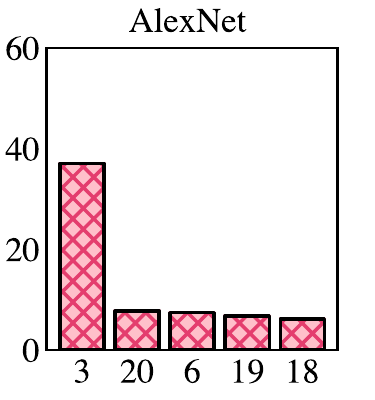}
\includegraphics[width=0.115\textwidth]{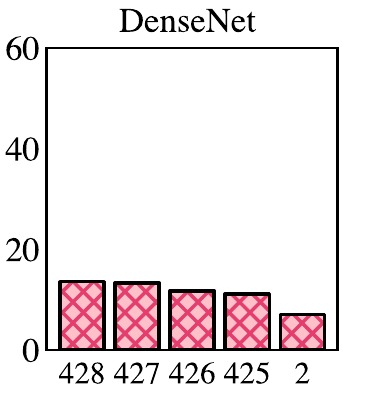}
\includegraphics[width=0.115\textwidth]{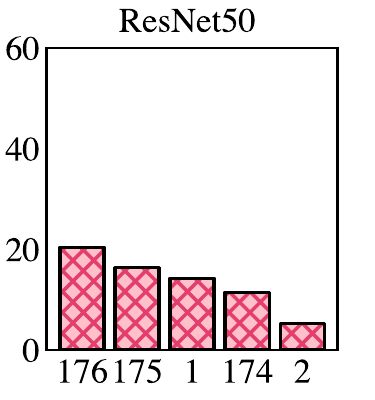}
\includegraphics[width=0.115\textwidth]{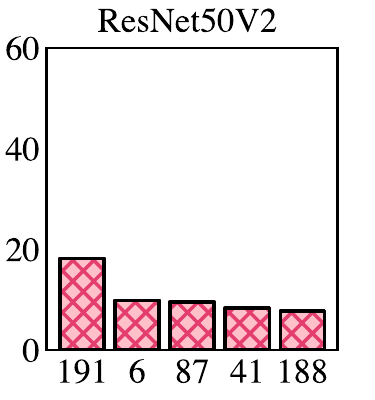}
\includegraphics[width=0.115\textwidth]{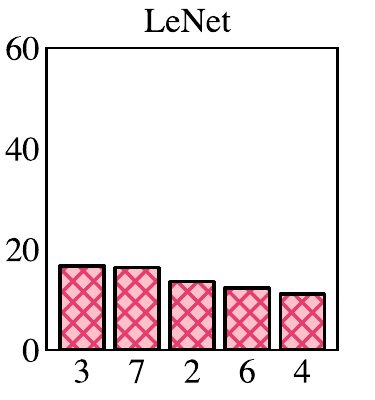}
\caption{Percentage of performance efficient partitioning points for the DNNs when there is an edge CPU stress of 90\% for different memory stress on the edge and network data transfer rates between the edge and the cloud on all experimental platforms $P_1$, $P_2$, $P_3$ and $P_4$. X-axis shows the partitioning point $n$, which is the $n^{th}$ layer of the DNN where the DNN is partitioned.}
\label{fig:CPUstress-maxstress}
\end{figure}

\begin{figure}[t]
\centering
\includegraphics[width=0.115\textwidth]{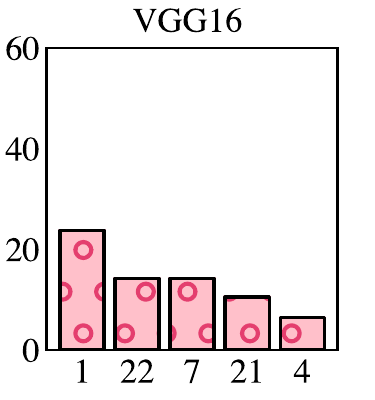}
\includegraphics[width=0.115\textwidth]{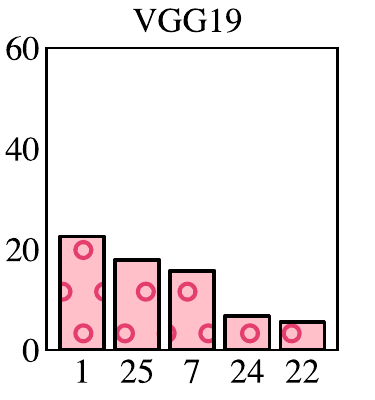}
\includegraphics[width=0.115\textwidth]{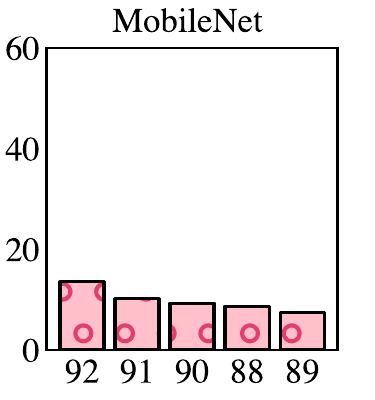}
\includegraphics[width=0.115\textwidth]{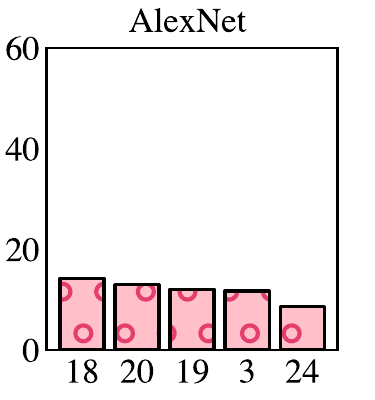}
\includegraphics[width=0.115\textwidth]{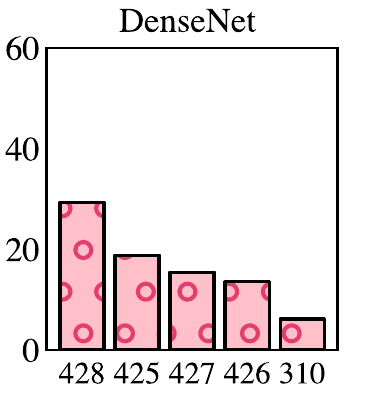}
\includegraphics[width=0.115\textwidth]{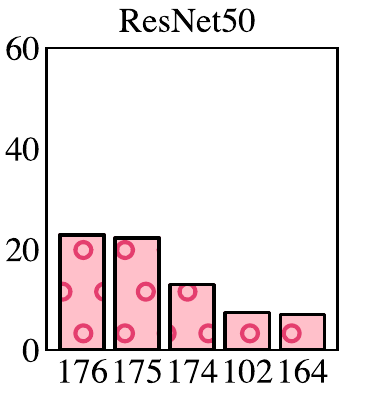}
\includegraphics[width=0.115\textwidth]{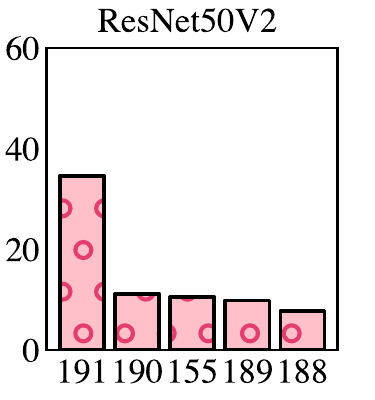}
\includegraphics[width=0.115\textwidth]{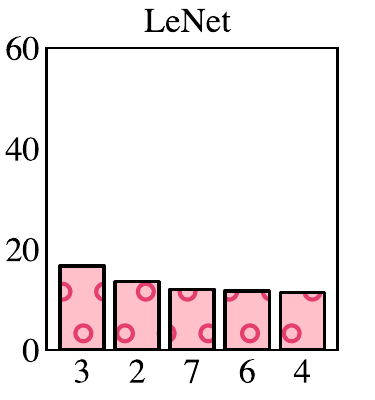}
\caption{Percentage of performance efficient partitioning points for the DNNs when there is 0\% edge CPU stress for different memory stress on the edge and network data transfer rates between the edge and the cloud on all experimental platforms $P_1$, $P_2$, $P_3$ and $P_4$. X-axis shows the partitioning point $n$, which is the $n^{th}$ layer of the DNN where the DNN is partitioned.}
\label{fig:CPUstress-minstress}
\end{figure}

Figure~\ref{fig:CPUstress-maxstress} and Figure~\ref{fig:CPUstress-minstress} show the percentage of the top five performance efficient partitioning points for the DNNs when there is a CPU stress of 90\% and of 0\% on the edge, respectively, when there is different memory stress at the edge and data transfer rates between the edge and the cloud.
The general observation is that partitions with more layers on the cloud are appropriate when CPU stress is maximum. For example, consider AlexNet. The two prominent partitioning layers when CPU stress is at a minimum are 18 and 20. However, when CPU stress is at 90\% the optimal partitioning point is layer 3. \revision{  }
A few observations from the results are that: (i) for VGG16 layer 1 and layer 22 are usually the prominent performance efficient partitioning points when CPU stress is 0\% meanwhile layer 4 becomes a prominent partitioning point when CPU stress is 100\%, (ii) AlexNet becomes less sensitive to increased CPU stress as Layer 3 becomes a prominent partitioning point, (iii) the optimal partitioning point for DenseNet is Layer 428 for both minimum and maximum CPU stress, even as network and memory performance vary.

\begin{figure}[t]
\centering
\includegraphics[width=0.115\textwidth]{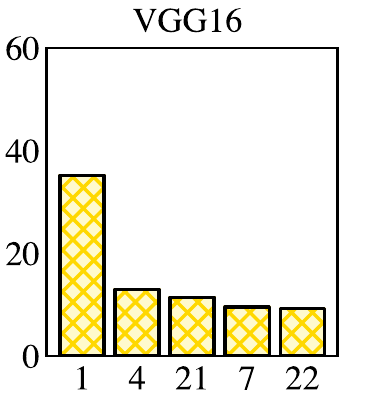}
\includegraphics[width=0.115\textwidth]{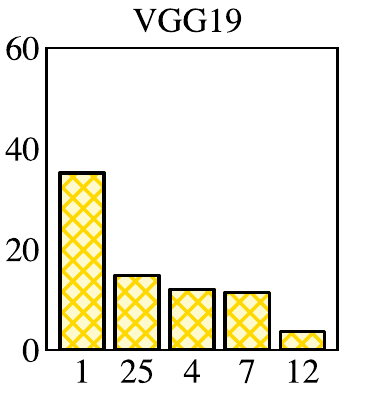}
\includegraphics[width=0.115\textwidth]{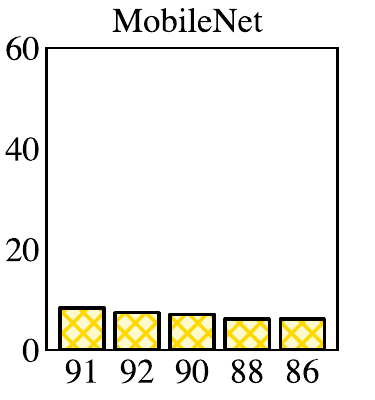}
\includegraphics[width=0.115\textwidth]{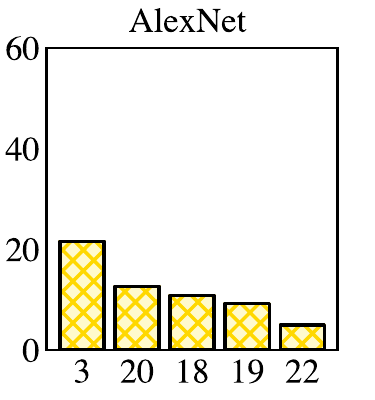}
\includegraphics[width=0.115\textwidth]{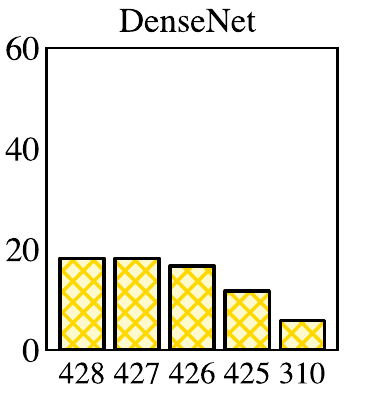}
\includegraphics[width=0.115\textwidth]{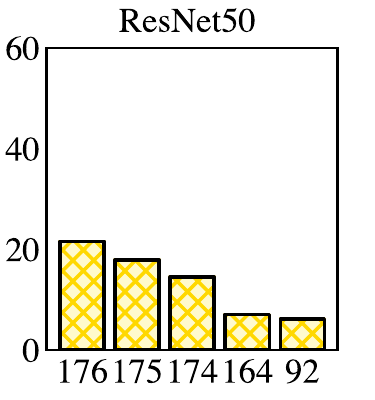}
\includegraphics[width=0.115\textwidth]{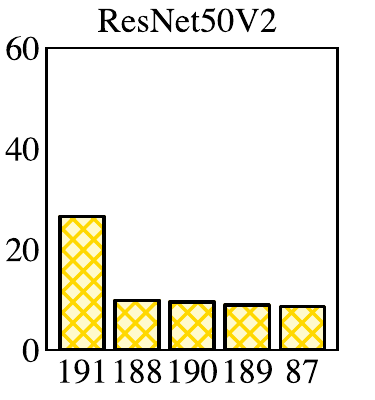}
\includegraphics[width=0.115\textwidth]{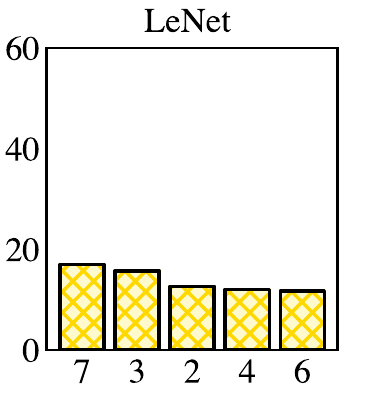}
\caption{Percentage of performance efficient partitioning points for the DNNs for a maximum memory stress of 90\% at the edge for different CPU stress on the edge and network data transfer rate on all experimental platforms $P_1$, $P_2$, $P_3$ and $P_4$. X-axis shows the partitioning point $n$, which is the $n^{th}$ layer of the DNN where the DNN is partitioned.}
\label{fig:memorystress-maxstress}
\end{figure}

\begin{figure}[t]
\centering
\includegraphics[width=0.115\textwidth]{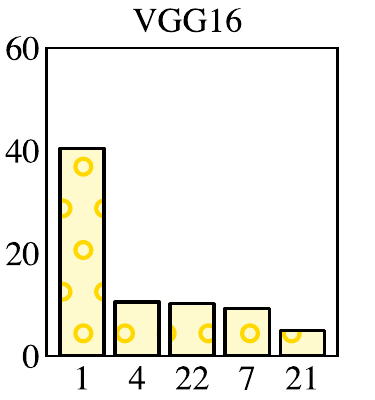}
\includegraphics[width=0.115\textwidth]{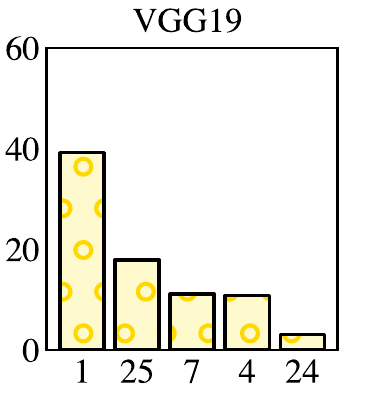}
\includegraphics[width=0.115\textwidth]{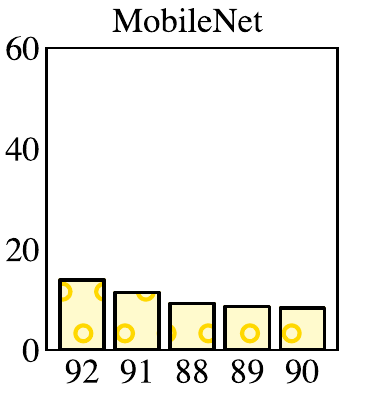}
\includegraphics[width=0.115\textwidth]{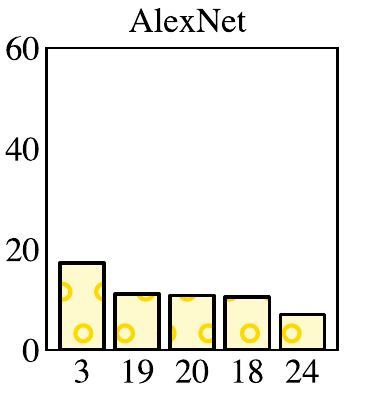}
\includegraphics[width=0.115\textwidth]{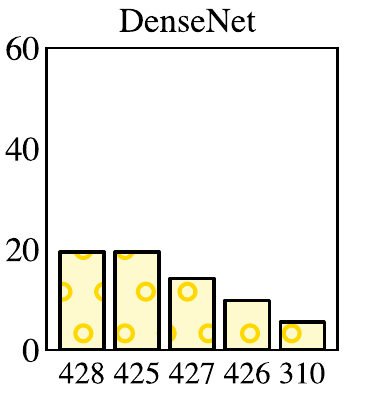}
\includegraphics[width=0.115\textwidth]{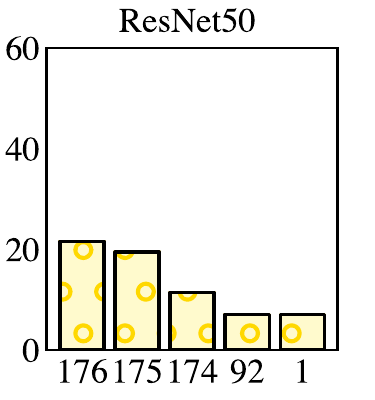}
\includegraphics[width=0.115\textwidth]{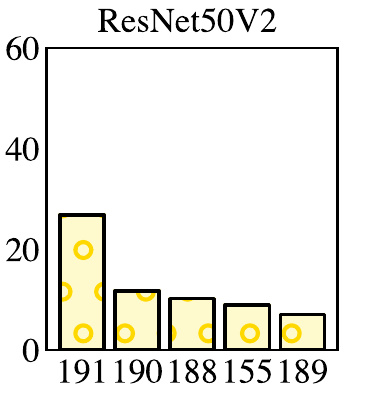}
\includegraphics[width=0.115\textwidth]{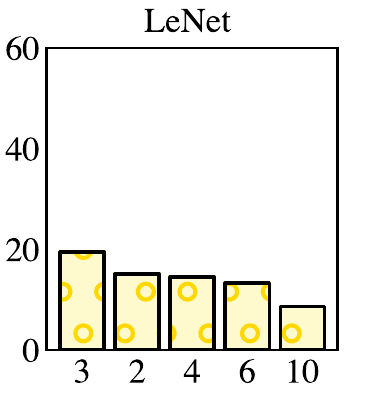}
\caption{Percentage of performance efficient partitioning points for the DNNs for a minimum memory stress of 0\% at the edge for different CPU stress on the edge and network data transfer rate on all experimental platforms $P_1$, $P_2$, $P_3$ and $P_4$. X-axis shows the partitioning point $n$, which is the $n^{th}$ layer of the DNN where the DNN is partitioned.}
\label{fig:memorystress-minstress}
\end{figure}

Figure~\ref{fig:memorystress-maxstress} and Figure~\ref{fig:memorystress-minstress} show the percentage of the top five performance efficient partitioning points for the DNNs when there is a memory stress of 90\% and 0\% on the edge, respectively, for different CPU stress at the edge and data transfer rates between the edge and the cloud. The general observation is that there are fewer changes to the optimal partitioning points. Memory stress does not impact the partitioning profile significantly. 

\begin{figure}[t]
\centering
\includegraphics[width=0.115\textwidth]{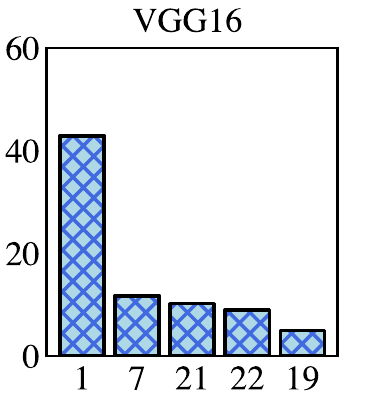}
\includegraphics[width=0.115\textwidth]{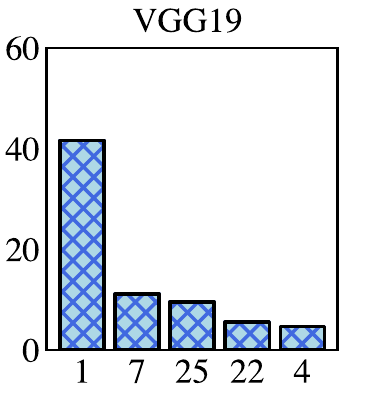}
\includegraphics[width=0.115\textwidth]{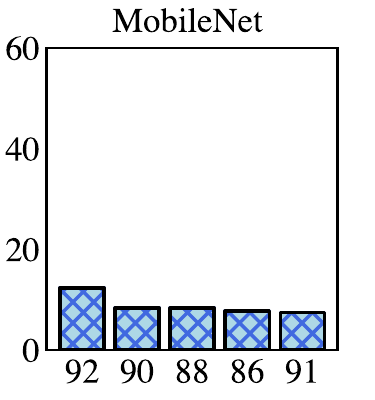}
\includegraphics[width=0.115\textwidth]{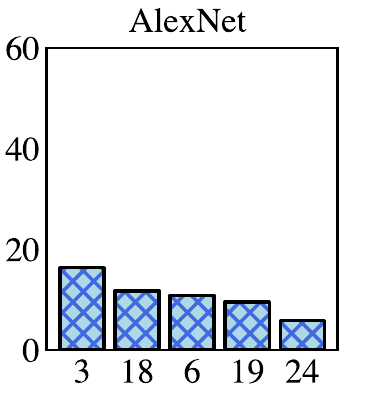}
\includegraphics[width=0.115\textwidth]{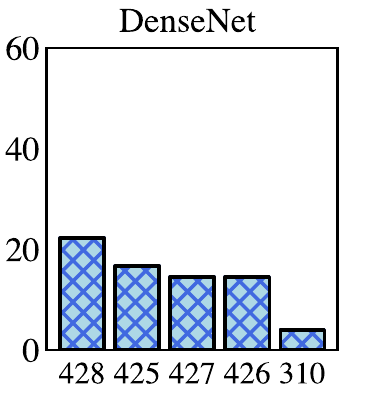}
\includegraphics[width=0.115\textwidth]{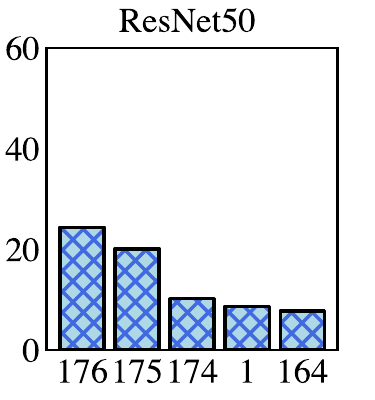}
\includegraphics[width=0.115\textwidth]{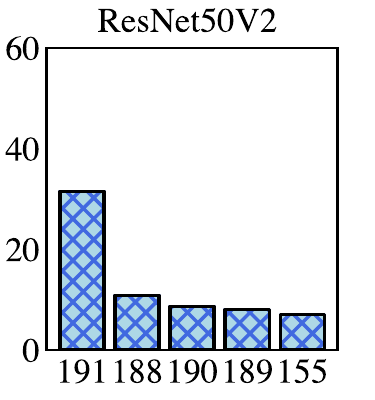}
\includegraphics[width=0.115\textwidth]{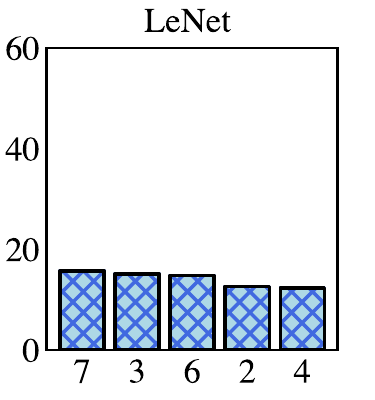}
\caption{Percentage of performance efficient partitioning points for the DNNs when the network data transfer rate between the edge and the cloud is 50Mb/s under different CPU and memory stress on the edge on all experimental platforms $P_1$, $P_2$, $P_3$ and $P_4$. X-axis shows the partitioning point $n$, which is the $n^{th}$ layer of the DNN where the DNN is partitioned.}
\label{fig:networkstress-maxavailable}
\end{figure}

\begin{figure}[t]
\centering
\includegraphics[width=0.115\textwidth]{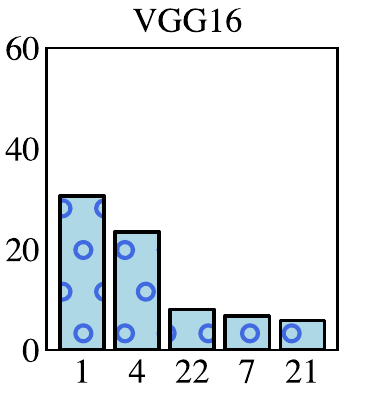}
\includegraphics[width=0.115\textwidth]{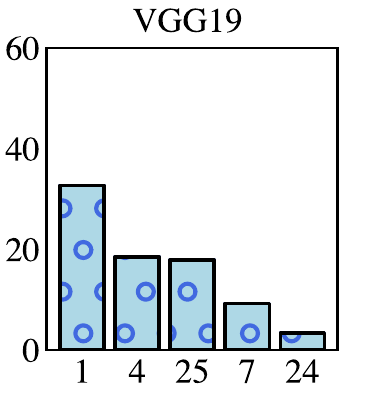}
\includegraphics[width=0.115\textwidth]{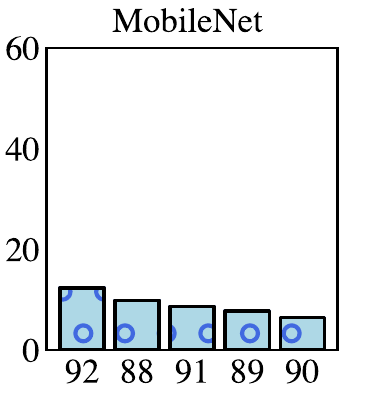}
\includegraphics[width=0.115\textwidth]{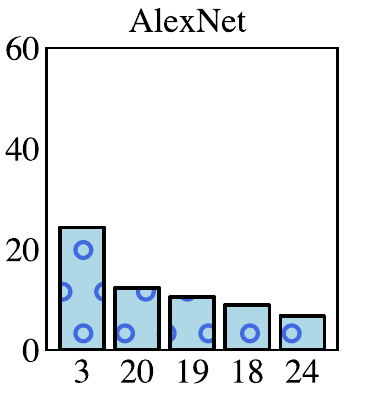}
\includegraphics[width=0.115\textwidth]{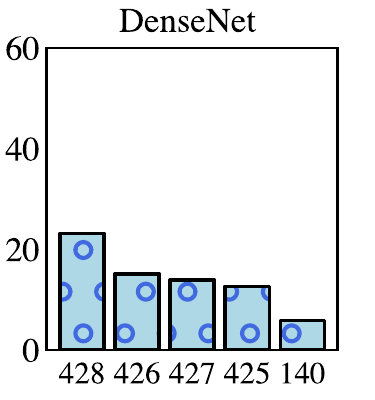}
\includegraphics[width=0.115\textwidth]{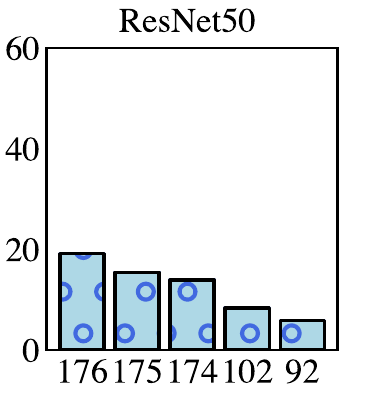}
\includegraphics[width=0.115\textwidth]{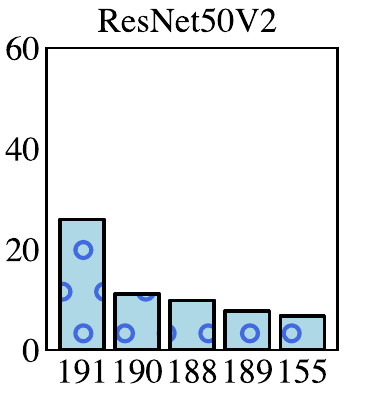}
\includegraphics[width=0.115\textwidth]{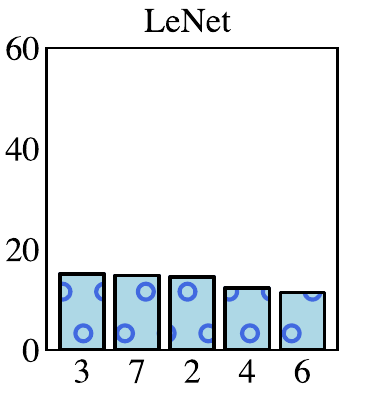}
\caption{Percentage of performance efficient partitioning points for the DNNs when the network data transfer rate between the edge and the cloud is 10Mb/s under different CPU and memory stress on the edge on all experimental platforms $P_1$, $P_2$, $P_3$ and $P_4$. X-axis shows the partitioning point $n$, which is the $n^{th}$ layer of the DNN where the DNN is partitioned.}
\label{fig:networkstress-minavailable}
\end{figure}

Figure~\ref{fig:networkstress-maxavailable} and Figure~\ref{fig:networkstress-minavailable} show the percentage of performance efficient partitioning points for the DNNs when the network data transfer rate between the edge and the cloud is 50Mb/s and 10Mb/s under different CPU and memory stress levels on the edge. The graphs highlight that DNNs are sensitive to a combination of operational conditions. Although individual operational conditions, such as CPU or memory stress may not affect DNNs substantially, the combination of operational conditions makes a case for adaptive DNNs.

\begin{figure}[t]
\centering
\includegraphics[width=0.115\textwidth]{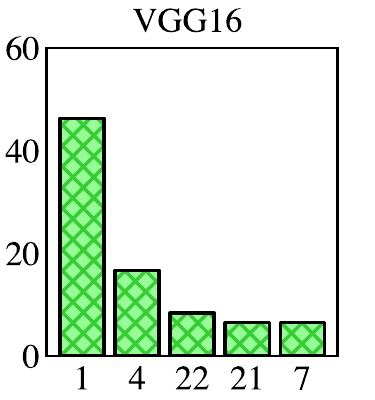}
\includegraphics[width=0.115\textwidth]{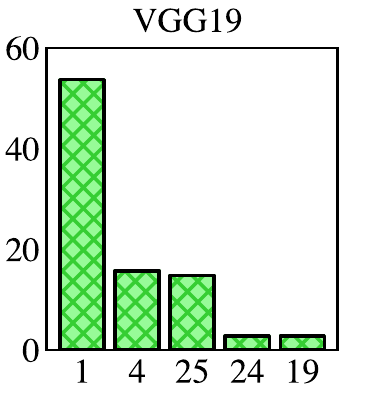}
\includegraphics[width=0.115\textwidth]{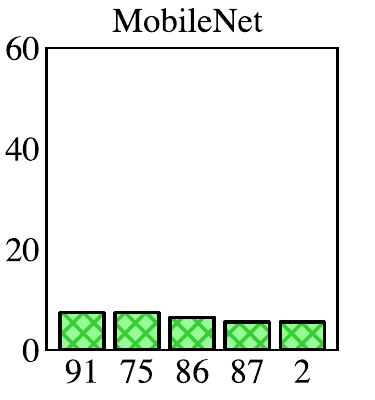}
\includegraphics[width=0.115\textwidth]{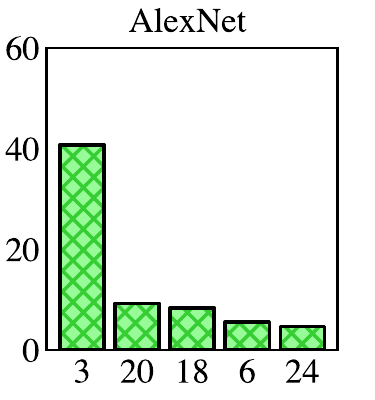}
\includegraphics[width=0.115\textwidth]{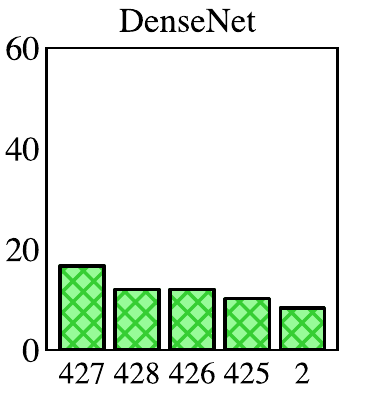}
\includegraphics[width=0.115\textwidth]{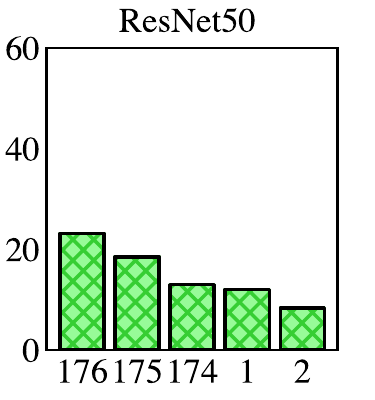}
\includegraphics[width=0.115\textwidth]{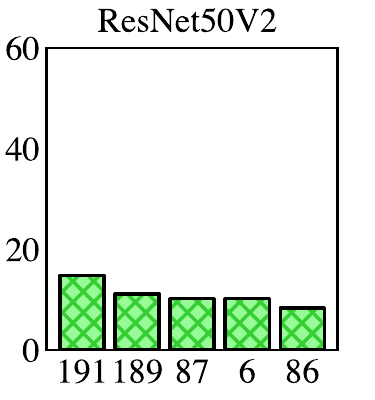}
\includegraphics[width=0.115\textwidth]{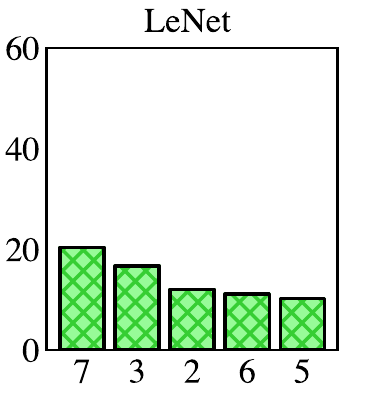}
\caption{Percentage of performance efficient partitioning points for the DNNs when there is maximum CPU and memory stress for different network data transfer rates on all experimental platforms $P_1$, $P_2$, $P_3$ and $P_4$. X-axis shows the partitioning point $n$, which is the $n^{th}$ layer of the DNN where the DNN is partitioned.}
\label{fig:comb-01}
\end{figure}

\begin{figure}[t]
\centering
\includegraphics[width=0.115\textwidth]{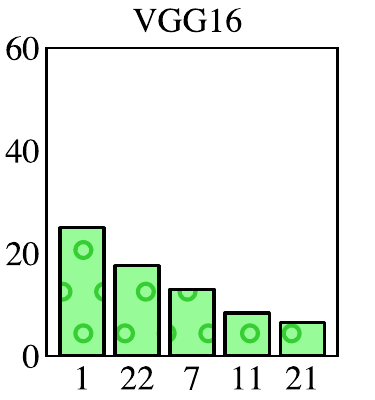}
\includegraphics[width=0.115\textwidth]{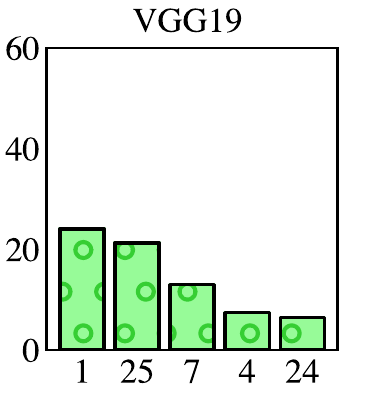}
\includegraphics[width=0.115\textwidth]{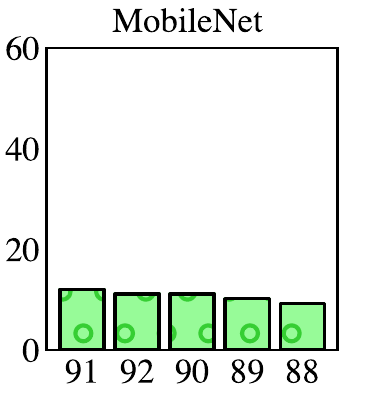}
\includegraphics[width=0.115\textwidth]{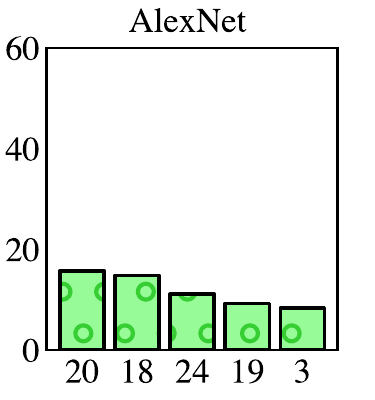}
\includegraphics[width=0.115\textwidth]{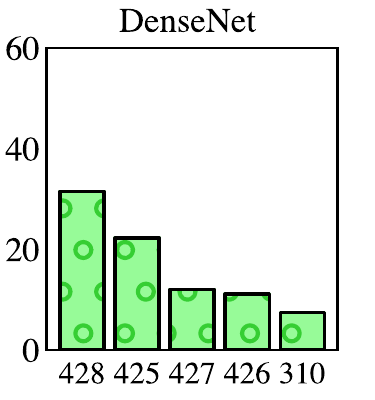}
\includegraphics[width=0.115\textwidth]{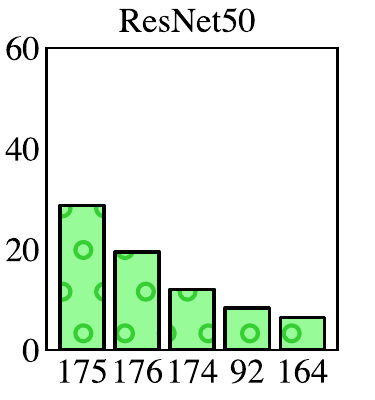}
\includegraphics[width=0.115\textwidth]{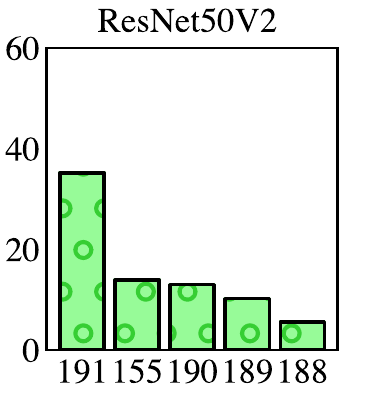}
\includegraphics[width=0.115\textwidth]{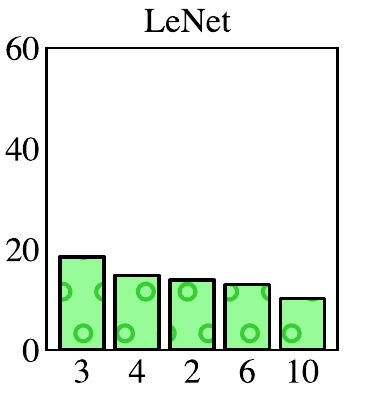}
\caption{Percentage of performance efficient partitioning points for the DNNs when there is no CPU and memory stress for different network data transfer rates on all experimental platforms $P_1$, $P_2$, $P_3$ and $P_4$. X-axis shows the partitioning point $n$, which is the $n^{th}$ layer of the DNN where the DNN is partitioned.}
\label{fig:comb-02}
\end{figure}

Figure~\ref{fig:comb-01} and Figure~\ref{fig:comb-02} show the percentage of performance efficient partitioning points for the DNNs for different network data transfer rates when there is maximum and minimum CPU and memory stress respectively. The general trend as previously observed is that DNNs are sensitive to a combination of operational conditions. However, when both figures are compared, VGG16, VGG19 and AlexNet tend to have a more dominant partitioning point when the CPU and memory stress is maximum. In other words, these DNNs are less sensitive to adaptivity under maximum CPU and memory stress. 
These observations are in line with those made previously.


\textit{Summary}: The experimental results have highlighted that the DNNs considered are sensitive to operational conditions and therefore are amenable to repartitioning. However, it is observed that the DNNs are more sensitive to network data transfer than CPU or memory stress when considered individually. A performance gain is observed when DNNs are repartitioned. A case for DNN adaptivity is stronger when a combination of operational conditions are considered. Although DNNs are sensitive to underlying hardware architectures as seen in Table~\ref{tab:sensitivity-compare-individual}, a specific pattern was not noted.  

\section{Related Work}
\label{sec:relatedwork}
The concept of adaptivity in DNNs has not been presented as a singular concept in the existing literature, rather is a reference to multi-faceted aspects of DNN execution. In this section, the presentation of two classes of adaptive DNNs is highlighted. The first is adaptivity in the context of DNNs that natively execute on a device or a server is considered. The second is adaptivity in the context of DNN partitioning. 

Adaptivity in the context of DNNs that natively execute on a device or a server is considered in three different ways. Firstly, DNN adaptivity for executing pre-trained models natively on a resource has been reported, for example in the acoustic context by taking into account the speaker or environment~\cite{acoustic-01}. Typically, the DNNs are fine-tuned to provide a higher quality result, primarily measured by accuracy~\cite{acoustic-02}. Three types of adaptive approaches are considered. The first is input feature transformation, the second is direct adaptation by transforming DNN parameters, and finally using the auxiliary context features (noise estimates which are provided as input to the DNN)~\cite{acoustic-01}. In the above, adaptivity is a reference to taking the acoustic environment or user into account for maximizing the performance of the DNN. 

Secondly, adaptivity is considered as choosing an appropriate DNN model for inference from a portfolio of pre-trained DNN models~\cite{adaptive-03}. For this a learning approach is used to estimate the optimal DNN model that will maximize the estimation accuracy for a particular type of input.

Thirdly, adaptivity is investigated in the context of model compression to execute DNN models on resource constrained devices~\cite{adaptive-04}. In this approach, the weight matrices of fully connected and convolutional neural networks are compressed without losing significant accuracy. The approach does not randomly drop components of the weight matrix, rather a more disciplined approach using singular value decomposition is used to prune fully-connected structures, which is what is referred to as being adaptive.
 
Adaptivity in the context of DNN partitioning is also presented in the literature. With the emergence of edge computing, DNN partitioning and distribution across devices, edge and cloud resources for inference has become an important avenue of research. Literature would suggest that adaptivity in the context of partitioning and distributing DNNs across multiple resources is not only about improving accuracy, but about optimizing end-to-end latencies. In other words, adaptivity is understood as making DNN partitions suitable for the operational context. 

There are numerous DNN partitioning methods presented in the literature.
These methods include, estimation-based techniques~\cite{neurosurgeon, deepwear, musicalchair, couper}, structural modification-based techniques~\cite{deepthings, modnn}, and measurement-based techniques~\cite{lavea}. These techniques identify an optimal partitioning point primarily based on the characteristics of the layers of a DNN and operational conditions, such as resource utilization or network conditions.

The above referenced research on DNN partitioning focuses on the initial partitioning and deployment of the DNNs. They do not focus on whether and how DNNs are sensitive to operational conditions after the initial deployment. This is significantly important for use-cases that are latency-critical and operate in a transient environment. Consider for example, an autonomous car or a fleet of drones for which a DNN was partitioned and deployed. The initial partition configuration may not be an optimal configuration or may even be obsolete with regard to performance given changes in the operational environment. This would require the DNN to be repartitioned for a new configuration.  

Currently, the sensitivity of DNNs to operational conditions of the edge resource (resource utilization) and the network state between the edge and the cloud for repartitioning is understood in a limited way. This is because the above referenced research has considered the initial deployment of a partitioned DNN. 

There is recent research that focuses on adaptivity for DNN partitioning~\cite{adaptive-05}. However, the main consideration is performance-awareness for which metrics, such as per layer latency, are used to inform DNN partitioning. A combination of an estimation-based approach is used for finding the partitioning point. An early exit strategy is employed for further optimizing performance. The research is pursued in the context of initial partitioning and the effect of evolving operational conditions are not explored. 

The impact of network conditions for DNN repartitioning has been considered~\cite{adaptive-06}. The network is assumed to have two states - lightly and heavily loaded states. However, how sensitive are DNNs to operational conditions (both resource specific and network specific) is not considered. 

Similarly, there is research that considers adaptivity in the context of partitioning output data from the layers using a compression-based technique, referred to as the compressed sparse row scheme~\cite{adaptive-07}. The output matrices of a layer are partitioned into dense and sparse partitions. This compression relies on network conditions and hence is referred to as being adaptive.

Privacy for adaptive partitioning of DNNs is considered by employing the Kullback-Leibler divergence between the intermediate layer outputs and the original input~\cite{adaptive-08}. A Lyapunov optimization framework is employed for partitioning the DNNs. 

Contrary to the above considerations, this paper sets out to investigate whether there is a case for adaptive DNNs in edge computing. The key questions that concern this paper in the context of adaptivity are whether DNNs are sensitive to operational conditions, and if so, how sensitive are they. In addition, the study explores the effect of individual and a combination of operational conditions on DNNs.

\section{Conclusions}
\label{sec:conclusions}
Performance-critical and privacy-sensitive applications benefit from edge computing by distributing selected services of an application closer to where data is generated. This allows applications to pre-process and selectively release data to make the overall application more responsive and privacy-aware. Deep Neural Networks (DNNs) are a class of applications that naturally lend themselves to distribution across the cloud and edge given that they are organized as a sequence of layers. The distribution of a DNN is achieved by partitioning it at a layer that would maximize its performance while taking operational conditions where the distributed DNN will be deployed into account (for example, CPU/memory stress on an edge resource or network data transfer rates between the edge and the cloud). 

There is limited understanding of how evolving operational conditions might affect the performance of a distributed DNN and whether a new partition is required to optimize the overall performance. If operational conditions affect DNNs, then they will need to be repartitioned (a new layer at which the DNN can be partitioned needs to be identified). A DNN that adapts to the operational conditions is defined in this paper as an `adaptive DNN'. 

This paper set out to investigate \textit{whether there is a case for adaptive DNNs in edge computing}. In doing so, the following three questions were considered: (i) Are DNNs sensitive to operational conditions? (ii) How sensitive are DNNs to operational conditions? (iii) Do individual or a combination of operational conditions equally impact DNNs? (iv) Is DNN partitioning sensitive to hardware architectures employed on the cloud/edge? 
To address the above an exploratory exercise was carried out by benchmarking 8 pre-trained production DNNs for different operational conditions, such as CPU/memory stress and network data transfer rates. The results presented were obtained from a cloud-edge lab-based experimental platform by analyzing nearly 8 million data points (the data and code is available for public download\footnote{\url{https://github.com/qub-blesson/AdaptiveDNN}}). The key observations are that DNNs are sensitive to both individual and a combination of operational conditions. When considering individual operational conditions, network conditions have a more substantial impact on DNN performance than CPU/memory stress on an edge node. Repartitioning can provide performance gains for certain DNNs considered. Operational conditions in combination have a more significant impact on DNN repartitioning than individual operational conditions. DNN partitioning is sensitive to the underlying hardware architecture, but a pattern was not noted. This paper concludes that there is a case for adaptive DNNs at the edge.

One of the limitations of the current exploration is that the case for adaptivity assumes the execution of the DNNs on CPUs. There is a compelling case in the real-world for using hardware accelerator platforms, such as GPUs, TPUs and ASICs. Distributed inference on such platforms, with applied stresses to investigate whether there is a case for adaptivity, will be considered in the future.
Since the number of combinations for varying stress in reality can be significantly higher than those considered in this paper, a learning approach may be useful to estimate best partition points. Such an approach or the measurement-based approach presented in this paper can be incorporated in an orchestration management system that deploys distributed DNNs. In addition, the implementation of adaptive DNNs for real applications and exploring domain-specific performance benefits will be considered. 

\section*{Acknowledgment}
Dr Blesson Varghese is supported by a Royal Society Short Industry Fellowship to British Telecommunications plc, UK, and by funds from Rakuten Mobile, Japan. 

\balance
\bibliographystyle{IEEEtran} 
\bibliography{references}

\end{document}